\documentclass[twocolumn]{aastex62}

\usepackage{amsmath}
\usepackage{rotating}
\graphicspath{{./}{figures/}}

\newcommand{\bvec}[1]{{\ensuremath{\bf{#1}}}}
\newcommand{\thetabv}{{\ensuremath{\theta_{BV}}}}

\begin{document}

\title{Ion Scale Electromagnetic Waves in the Inner Heliosphere}

\begin{abstract}
Understanding the physical processes in the solar wind and corona which actively contribute to heating, acceleration, and dissipation is a primary objective of NASA's Parker Solar Probe (PSP) mission. Observations of coherent electromagnetic waves at ion scales suggests that linear cyclotron resonance and non-linear processes are dynamically relevant in the inner heliosphere. A wavelet-based statistical study of coherent waves in the first perihelion encounter of PSP demonstrates the presence of transverse electromagnetic waves at ion resonant scales which are observed in 30-50\% of radial field intervals. Average wave amplitudes of approximately 4 nT are measured, while the mean duration of wave events is of order 20 seconds; however long duration wave events can exist without interruption on hour-long timescales. Though ion scale waves are preferentially observed  during intervals with a radial mean magnetic field, we show that measurement constraints, associated with single spacecraft sampling of quasi-parallel waves superposed with anisotropic turbulence, render the measured quasi-parallel ion-wave spectrum unobservable when the mean magnetic field is oblique to the solar wind flow; these results imply that the occurrence of coherent ion-scale waves is not limited to a radial field configuration. The lack of strong radial scaling of characteristic wave amplitudes and duration suggests that the waves are generated {\em{in-situ}} through plasma instabilities. Additionally, observations of proton distribution functions indicate that temperature anisotropy may drive the observed ion-scale waves.

\end{abstract}
\author[0000-0002-4625-3332]{Trevor  A. Bowen}
\affil{Space Sciences Laboratory, University of California, Berkeley, CA 94720-7450, USA}
\correspondingauthor{Trevor A.~Bowen}
\email{tbowen@berkeley.edu}

\author{Alfred Mallet}
\affiliation{Space Sciences Laboratory, University of California, Berkeley, CA 94720-7450, USA}

\author[0000-0002-9954-4707]{Jia Huang}
\affiliation{Climate and Space Sciences and Engineering, University of Michigan, Ann Arbor, MI 48109, USA}

\author[0000-0001-6038-1923]{Kristopher G. Klein}
\affiliation{University of Arizona, Tucson, AZ}

\author[0000-0003-1191-1558]{David M. Malaspina}
\affil{Laboratory for Atmospheric and Space Physics, University of Colorado, Boulder, CO 80303, USA}

\author[0000-0002-7728-0085]{Michael Stevens}
\affiliation{Smithsonian Astrophysical Observatory, Cambridge, MA 02138 USA}

\author[0000-0002-1989-3596]{Stuart D. Bale}
\affil{Space Sciences Laboratory, University of California, Berkeley, CA 94720-7450, USA}
\affil{Physics Department, University of California, Berkeley, CA 94720-7300, USA}
\affil{The Blackett Laboratory, Imperial College London, London, SW7 2AZ, UK}
\affil{School of Physics and Astronomy, Queen Mary University of London, London E1 4NS, UK}

\author[0000-0002-0675-7907]{J. W. Bonnell}
\affil{Space Sciences Laboratory, University of California, Berkeley, CA 94720-7450, USA}

\author[0000-0002-3520-4041]{Anthony W. Case}
\affiliation{Smithsonian Astrophysical Observatory, Cambridge, MA 02138 USA.}

\author[0000-0003-4177-3328]{Benjamin D. G. Chandran}
\affil{Department of Physics \& Astronomy, University of New Hampshire, Durham, NH 03824, USA}
\affil{Space Science Center, University of New Hampshire, Durham, NH 03824, USA}

\author[0000-0002-0029-717X]{C. C. Chaston}
\affiliation{Space Sciences Laboratory, University of California, Berkeley, CA 94720-7450, USA}

\author[0000-0003-4529-3620]{Christopher H. K. Chen}
\affil{School of Physics and Astronomy, Queen Mary University of London, London E1 4NS, UK}

\author[0000-0002-4401-0943]{Thierry {Dudok de Wit}}
\affiliation{LPC2E, CNRS and University of Orl\'eans, Orl\'eans, France}

\author[0000-0003-0420-3633]{Keith Goetz}
\affiliation{School of Physics and Astronomy, University of Minnesota, Minneapolis, MN 55455, USA}

\author[0000-0002-6938-0166]{Peter R. Harvey}
\affiliation{Space Sciences Laboratory, University of California, Berkeley, CA 94720-7450, USA}

\author[0000-0003-1749-2665]{Gregory G. Howes}
\affil{Department of Physics and Astronomy, University of Iowa, Iowa City, IA 52242, USA}

\author[0000-0002-7077-930X]{J. C. Kasper}
\affiliation{Climate and Space Sciences and Engineering, University of Michigan, Ann Arbor, MI 48109, USA}
\affiliation{Smithsonian Astrophysical Observatory, Cambridge, MA 02138 USA}

\author[0000-0001-6095-2490]{Kelly E. Korreck}
\affiliation{Smithsonian Astrophysical Observatory, Cambridge, MA 02138 USA}

\author{Davin Larson}
\affiliation{Space Sciences Laboratory, University of California, Berkeley, CA 94720-7450, USA}

\author{Roberto Livi}
\affiliation{Space Sciences Laboratory, University of California, Berkeley, CA 94720-7450, USA}

\author[0000-0003-3112-4201]{Robert J. MacDowall}
\affiliation{Solar System Exploration Division, NASA/Goddard Space Flight Center, Greenbelt, MD, 20771}

\author{Michael D. McManus}
\affiliation{Space Sciences Laboratory, University of California, Berkeley, CA 94720-7450, USA}
\affiliation{Physics Department, University of California, Berkeley, CA 94720-7300, USA}

\author[0000-0002-1573-7457]{Marc Pulupa}
\affiliation{Space Sciences Laboratory, University of California, Berkeley, CA 94720-7450, USA}

\author{J. L. Verniero}
\affiliation{Space Sciences Laboratory, University of California, Berkeley, CA 94720-7450, USA}

\author[0000-0002-7287-5098]{Phyllis Whittlesey}
\affiliation{Space Sciences Laboratory, University of California, Berkeley, CA 94720-7450, USA}

\collaboration{(The PSP/FIELDS and PSP/SWEAP Teams)}

\section{Introduction}

Statistical studies of the fluctuating solar wind reveal an environment, reminiscent of hydrodynamic turbulence, where non-linear interactions cause a cascade of energy from large to small scales \citep{Iroshnikov1963,Kraichnan1965,Coleman1968}. While robust theories have been developed to explain the inertial range of magnetohydrodynamic turbulence, e.g. \cite{GoldreichSridhar1995, Boldyrev2006,Schekochihin2009}, it is known that the solar wind is populated with coherent structures co-existing with the turbulent background. The existence of coherent structures, which can take the form of phase-coherent discontinuities, e.g. current sheets, and shocks, which can exist at many scales, or scale confined coherent structures such as waves and solitary vortices, are frequently connected to dissipation and heating processes \citep{Sundkvist2005,Alexandrova2006,Alexandrova2008,Osman2012,DudokdeWit2013,Matthaeus2015,Lion2016,Mallet2019}. 

The collisionless nature of the solar wind suggests that wave-particle interactions are necessary for the dissipation of magnetized turbulence. The ion-cyclotron resonance, which enables coupling of electromagnetic fluctuations with ion gyromotion (e.g. \citealt{Stix1992}), may contribute to coronal heating through damping of Alfv\'{e}nic fluctuations at ion gyroscales \citep{HollwegJohnson1988,TuMarsch1997,Cranmer2000,Cranmer2014}. Measurements of ion temperature anisotropies in the upper corona by ultraviolet spectroscopy suggest large $T_\perp/T_\parallel$ consistent with heating through cyclotron damping \citep{Kohl1997,Kohl1998,Cranmer1999,Cranmer2000}. 

In the solar wind, observations of non-zero magnetic helicity at ion scales have been interpreted as evidence for active cyclotron damping of turbulent fluctuations which contribute to the dissipation and heating at kinetic scales \citep{Isenberg1990,Leamon1998,Woodham2018}. In addition to proton-cyclotron dynamics, the cyclotron resonance of doubly ionized helium ($\alpha$-particles) and heavy ions has been studied extensively \citep{IsenbergHollweg1983,Isenberg1984}. Using observations of protons and $\alpha$-particles \cite{Kasper2013} argue that temperature anisotropy observed at 1AU is consistent with {\em{in-situ}} cyclotron resonant heating; though alternative theories may produce consistent solutions with other dissipation mechanisms, e.g. stochastic heating \citep{Chandran2013}.

Observational studies of magnetic helicity of the solar wind by \cite{PodestaGary2011} and \cite{He2011} reveal circular polarized fluctuations near the proton gyroscale, $\rho=V_{\perp pth}/\Omega_p$, where $V_{\perp pth}$ is the proton thermal speed perpendicular to the mean magnetic field, $B_0$, and $\Omega_p=qB_0/m_p$. Both studies found opposite polarizations at $|\text{cos}\theta_{BV}|\approx1$  and $|\text{cos}\theta_{BV}|\approx 0$, where $\thetabv$ is the angle between the mean magnetic field and the solar wind flow direction. \cite{PodestaGary2011} suggest that the observations are consistent with parallel propagating ion-cyclotron wave (ICWs) or fast-magnetosonic/whistler (FM/W) waves and anisotropic kinetic Alf\'{e}nic turbulence. \cite{He2011} interpret the observations as the result of a parallel propagating ICW population and oblique FM/W population. 

\cite{PodestaGary2011b} demonstrate the generation of ICW waves propagating anti-sunward and electron resonant FM/W waves propagating sun-ward from through the respective Alf\'{e}n/ion cyclotron and parallel fire-hose instabilities; the authors further argue that the Doppler shift of sunward propagating electron waves causes both modes to appear ion-resonant in the spacecraft frame. \cite{Klein2014} reproduced the observations of  helicity by \cite{PodestaGary2011} and \cite{He2011} using a superposition of quasi-perpendicular Alfv\'enic turbulence with quasi-parallel propagating ion cyclotron waves (ICWs) and electron resonant whistler waves, concluding that measurements of helicity alone are not sufficient to distinguish the wave modes. Using a $k$-filtering technique \cite{RobertsLi2015} determined that observed wave-vectors were consistent with ion-resonant fluctuations rather than a FM/W mode. 

While observations of the spectrum parallel to the magnetic field ($k_\parallel$) at kinetic scales suggest the presence of background fluctuations with an intrinsic non-zero helicity, large-amplitude coherent ICWs have been observed in many space-plasma environments. Observations of ICWs near planetary bodies are commonly associated with pick up of ions from neutral atmospheres \citep{Russell1990,Kivelson1996,Brain2002,Delva2011}. {Using ISEE-3 magnetometer data, \cite{Tsurutani1994} observed elliptically polarized ion scale waves upstream of the Earth with several nT amplitudes on order the mean field strength. Though the authors conjectured that the waves result from pickup-ion interaction, they could not rule out generation through instabilities or a solar source. \citep{Murphy1995} noted that the presence of circularly polarized coherent ion-scale waves in Ulysses observations occurred preferentially with a radial field alignment; their observations, spanning roughly two years and taken far from planetary sources, led the authors to conclude that interstellar pickup ion interactions were the source of these waves. 

Using the twin STEREO spacecraft spacecraft, \cite{Jian2009} performed a statistical study of ion-waves at 1AU spanning two months, identifying both left and right hand circular polarizations in the spacecraft frame with amplitudes on order $~0.1$ nT. The two month interval studied in \cite{Jian2009} coincides with an orbital configuration which precludes planetary generation from Earth and sufficient orbital separation such that observations of ion scale waves in each satellite could be taken independent events; their results showed that the presence of ICWs preferentially coincides with intervals of radial field, in agreement with the results of \cite{Murphy1995}. Subsequent work has revealed the presence of ICWs in the inner heliosphere with both {\em{MESSENGER}} and {\em{Helios}} data, again with a preference for radial field alignment \cite{Jian2010}. \cite{Boardsen2015} performed a multi-year analysis of frequency wave storms using MESSENGER observations. The radial scaling of coherent cyclotron polarized waves was found to be $\sim r^{-3}$ consistent with a WKB-like propagation suggested by \cite{Hollweg1974}, leading the authors to argue for the propagation of ICWs from the inner heliosphere. \cite{Jian2014} performed a statistical study of one year of STEREO data, comparing the occurrence of long duration ion-wave events; their results attempt to rule out generation from interstellar pickup or transient solar wind features (shocks, coronal mass ejections), suggesting that the waves may grow through plasma instabilities.

 The relatively weak collisionality of the hot and diffuse solar wind allow for the persistence of non-Maxwellian velocity distribution functions, which are typically parameterized by unequal temperatures along and transverse to the local magnetic field direction and distinct particle populations drifting with respect to one another. Both temperature anisotropy and drifting particle populations are commonly measured in the inner heliosphere and at 1 AU \citep{Marsch1982,Cranmer1999,Marsch2012,Wilson2018, Kasper2017}.} These non-Maxwellian features are capable of generating coherent waves through quasi-linear processes \citep{Gary1993book,Gary2001,Kasper2002,Verscharen2016,Verscharen2019}. The unstable waves can be driven by resonant wave-particle interactions, generating a number of different wave modes depending on the local plasma conditions; see \cite{Gary1993book} for a classic reference on these unstable modes and \cite{Yoon2017} for a discussion of their quasi-linear evolution. A number of different kinds of non-equilibrium structures can drive ion-cyclotron waves (e.g. see Table 4 in \citealt{Verscharen2019} and references therein) including proton and minor ion temperature anisotropies with $T_\perp>T_\parallel$ and relatively drifting ion populations. The conditions for these resonant instabilities, which have their fastest growing modes at wave-vectors $k_\parallel \rho \approx 1$ and $k_\perp \ll k_\parallel$, arise frequently in the solar wind \citep{Klein2018}.

Several studies have provided observational support for wave generation through instabilities. \cite{Wicks2016} demonstrate correlations between plasma beam energy and coherent wave power, as well as the coincidence of unstable distributions with the presence of long duration events, interpreting the observations as generation of large coherent wave through an anisotropic core-beam instabilities. Separately, \cite{Woodham2019} and \cite{Zhao2019} demonstrate increased occurrence of coherent waves when temperature anisotropy threshold conditions are met.

This manuscript reports detailed measurements of ion scale waves observed by NASA's Parker Solar Probe (PSP) \citep{Bale2019}. The presence of coherent waves near ion-kinetic scales and the inherent connection of these waves with kinetic instabilities suggest that ion-cyclotron resonance may play a role in coronal heating heating and solar wind acceleration. Furthermore the ubiquity of transverse ion waves in the inner heliosphere, in comparison with 1AU observations, suggest that the these waves, and the processes which drive and damp them, will become more prevalent as PSP descends into the solar corona. Through studying the statistical {\em{in-situ}} signatures of coherent waves in the inner heliosphere with instrumentation provided by the PSP electromagnetic FIELDS and Solar Wind Electron, Alpha, Proton (SWEAP) investigations, we aim to demonstrate the importance of these fluctuations to plasma dynamics in the inner heliosphere \citep{Bale2016,Kasper2016}.

Section \ref{sec:2} outlines the data acquisition, processing, and coherent wave identification and extraction using a continuous wavelet transform. Section \ref{sec:3} discusses the statistical observations of the coherent wave events in the inner heliosphere. Section \ref{sec:4} provides an analysis of sampling biases due to single spacecraft measurements of a multi-component wave-vector spectra consisting of anisotropic turbulence and parallel propagating waves. Section \ref{sec:5} compares the occurrence of wave events with estimates of proton core temperature anisotropy made through integrated 1D measurements. Section \ref{sec:6} compares the incidence of electrostatic waves with observations of ion-scale waves.

\section{Data Processing and Wavelet Analysis}\label{sec:2}

Data are obtained from the first PSP perihelion encounter (E1) between Oct 31- Nov 11, 2018. Measurements of the solar wind plasma are made by the Solar Wind Electrons Alphas and Protons (PSP/SWEAP) investigation \citep{Kasper2016,Fox2016}. Moments of ion distribution functions measured by the Solar Probe Cup (SPC) are used in evaluating temperature, $T_p$, solar wind velocity, $\bvec{V}_{sw}$, and density, $n_p$, over the first perihelion \citep{Case2019}.

The FIELDS instrument suite provides {\em{in-situ}} measurements of electromagnetic fluctuations on PSP \citep{Bale2016,Fox2016}. Magnetic field measurements are taken from the FIELDS flux-gate magnetometer (MAG). During E1, sample rates for the MAG vary between 73.242 Sa/sec and 292.969 Sa/sec. Data are down-sampled to a uniform rate of 36.621 Sa/sec, enabling uniform sampling over the first encounter while reducing computational loads and keeping the Doppler shifted ion scales within the bandwidth. A digital anti-aliasing filter is applied to remove power above the down-sampled Nyquist frequency ($\sim$18.32 Hz). SPC plasma data are typically sampled at a lower cadence than the MAG (roughly 0.25 - 1 sec). Plasma moments ($T_p$, $n_p$ and $V_{sw}$) are interpolated on to the 36.621 Sa/sec magnetic field time base. For each time step, the median value of the neighboring 512 samples (approximately 14 sec) is taken to remove impulsive noise and non-finite measurements. Additionally, using integrated observations of 1-D distribution functions from SWEAP/SPC, \cite{Huang2019} construct 3-D distribution functions with order 10-second cadence, providing estimates of the anisotropic perpendicular and parallel proton thermal velocities $V_{\perp th }$ and $V_{\parallel th }$.

The FIELDS magnetometer suite is sensitive to narrow-band coherent noise originating from the spacecraft reaction wheels. In order to minimize effects of the reaction wheels, which may contaminate magnetic field measurements at ion-scales, time dependent wheel rotation rates are identified from spacecraft housekeeping data. Power in the contaminated frequencies is attenuated in the Fourier domain, removing the polarized narrow-band spacecraft noise from the magnetic field data which can be mistaken as coherent plasma wave power. Appendix \ref{appendix} outlines the data processing technique and its impact on the results of the study.

\begin{figure}
    \centering
    \includegraphics[width=\columnwidth]{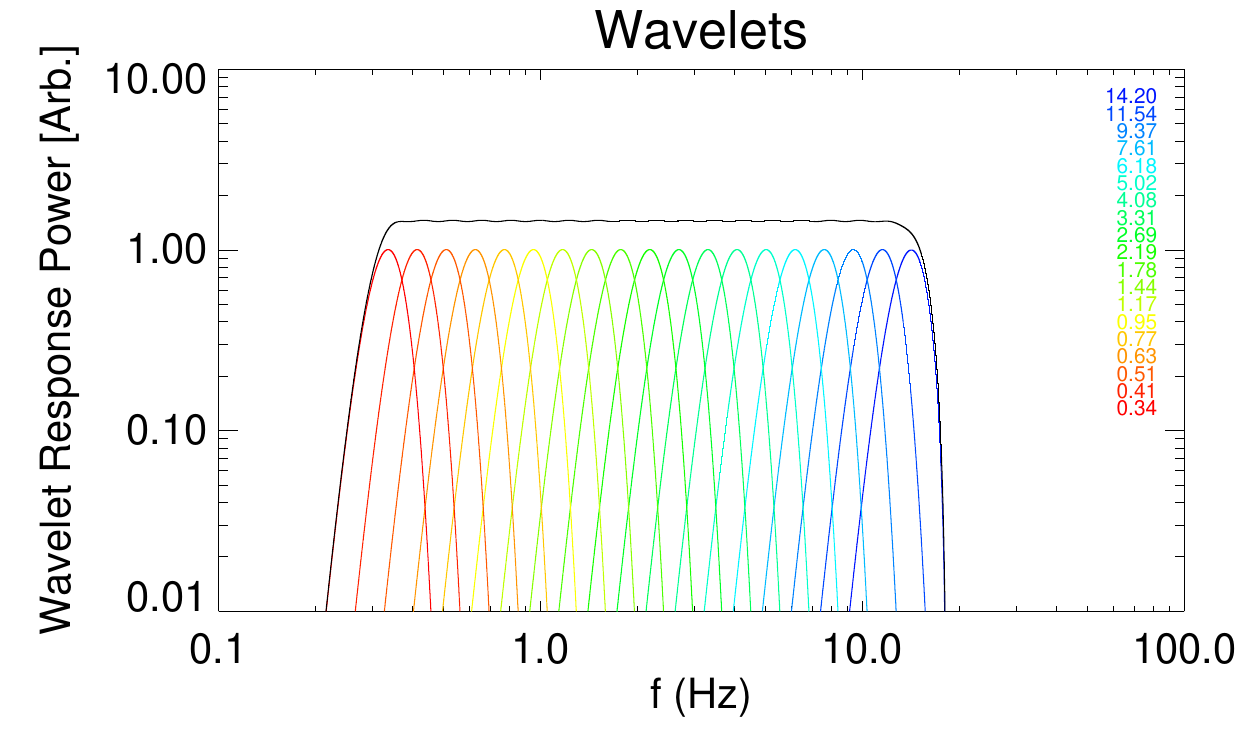}
    \caption{A set of 19 unnormalized Morlet wavelets (Gaussian modulated complex-exponential) with logarithmic spacing between $\sim$0.34 and 14.20 Hz. The black line shows the integrated response of the set of wavelets over the down-sampled ($36.621$ Sa/Sec) MAG bandwidth is uniform.}
    \label{fig:1}
\end{figure}

A wavelet transform is performed for each full day of down-sampled magnetic field data to obtain a time-frequency decomposition of the observations. The continuous wavelet transform is given as the convolution of a time-series 
with a set of scaled wavelets $\psi(s,\tau)$ normalized to unit energy at each scale:
\begin{equation}
    W(s,t)=\sum_{i=0}^{N-1} \psi\left(\frac{t_i-\tau}{s}\right)B(t_i)\\
   \end{equation}
   where the Morlet wavelet, given (in un-normalized form) as
\begin{equation}
\psi (\tau)= \pi^{1/4}e^{-i\omega_0\tau }e^\frac{-\tau^2}{2} \label{eq:wavelet},
\end{equation}

with non-dimensional time and frequency parameters $\tau$ and $\omega_0$, is used. \citep{Farge1992,TorrenceCompo,DudokdeWit2013}. The continuous wavelet transform of each component of the magnetic field time-series is taken using $\omega_0=6$, with the relationship between wavelet scale and spacecraft frequency approximated as $f\approx\frac{\omega_0}{2\pi s}f_s.$ A granular set of 19 scales ranging between $\approx0$.34 Hz and $\approx$ 14.2 Hz is used to capture the dynamics of coherent transverse waves near ion scales. Figure \ref{fig:1} shows the wavelet transform scales with the integrated response, which is nearly uniform over the considered range of frequencies. At each wavelet scale, the local measurement of the mean magnetic field is computed corresponding to the Gaussian envelope of each wavelet scale,

\begin{equation}
    B_0(s,t)=\sum_{i=0}^{N-1} \left|\psi(\frac{t_i-\tau}{s})\right|B(t_i)\label{eq:meanb}
   \end{equation}
   where $\left|\psi(\tau)\right|=\frac{A_s}{\pi^{1/4}}e^{-{\frac{\tau}{2}}^2}$ is normalized to unit energy \citep{Horbury2008, Podesta2009}.

  The vector wavelet transform in spacecraft coordinates $\vec{W}$ is then rotated into a field aligned coordinate system $\hat{W}=(\hat{B}_{\perp1},\hat{B}_{\perp2},\hat{B}_0)$. The first perpendicular unit vector, $\hat{B}_{\perp1}$, is computed as the cross product of the maximum variance direction of the mean field with the local mean field direction. The second unit vector, $\hat{B}_{\perp2}$, ensures closure of a right handed coordinate system $\hat{B}_{\perp1} \times \hat{B}_{\perp2}=\hat{B}_0$. To simplify notation, we use $B_{\perp1}(f,t)$ and $B_{\perp2}(f,t)$ to represent the complex valued wavelet transform of the magnetic field perpendicular to the mean direction and $B_\parallel$ as the complex valued wavelet transform parallel to the mean field.
 
 At each wavelet scale (frequency), quantities analogous to the Stokes parameters, are defined between the perpendicular components of the wavelet transform
\begin{eqnarray}
    S_0(f,t)=B_{\perp1}^2+B_{\perp2}^2\\
    S_1(f,t)=B_{\perp1}^2-B_{\perp2}^2\\
    S_2(f,t)=2\text{Re}(B_{\perp1}B_{\perp2}^*)\\ \label{eq:s2}
    S_3(f,t)=-2\text{Im}(B_{\perp1}B_{\perp2}^*) \label{eq:s3}
\end{eqnarray}

The quantity $S_3$ is associated with the magnetic helicity and circular polarization of the vector components $B_{\perp1}$ and  $B_{\perp2}$ along $B_0$ \citep{MatthaeusGoldstein1982,Narita2009,HowesQuataert2010}.

For a right handed coordinate system the vector $(\text{cos}(\phi),\text{sin}(\phi),B_0)$ with $B_0>0$, the phase convention of the Morlet wavelet in Equation \ref{eq:wavelet}, with $S_3$ defined in Equation \ref{eq:s3}, returns a normalized value $S_3/S_0=-1$. Physically, this polarization corresponds to the oscillation direction of electrons around the mean field in a stationary frame. The polarization corresponding to ion gyro-motion is associated with $S_3/S_0=1$. The normalized fraction of circularly polarized power, $\sigma(f,t)$ is defined as a time average

\begin{equation}
    \sigma(f,t)=\langle S_3\rangle/\langle S_0\rangle,
\end{equation}
over two e-foldings of the Gaussian envelope of the Morlet wavelet, which reduces fluctuations in $S_3$ associated with stochastic turbulent fluctuations with an instantaneous polarization.

Instead of relying on a pseudo-vector convention of  ``left'' and ``right'' handedness to describe helical fluctuations, we prefer the terms ion-resonant and electron-resonant polarization which are unambiguous vectors defined relative to $B_0$ and are intrinsic properties of the cold-plasma dispersion for parallel propagating electromagnetic waves: e.g. ICW resonate with ion gyromotion relative to $B_0$ \citep{Stix1992,Gary1993book}. Magnetic helicity is defined relative to the wave vector $\bvec{k}$ such that, e.g. an ion-resonant wave, has different values of helicity depending on its propagation direction \citep{Narita2009}. Due to single point measurement effects $\bvec{k}$ in the solar wind frame is typically not directly determined and the measured value of $S_3$ corresponds to the reduced magnetic helicity observed in the spacecraft frame \cite{MatthaeusGoldstein1982,HowesQuataert2010}. As noted by many authors, this degeneracy precludes a determination of the intrinsic plasma frame polarization of helicical fluctuations. In contrast the spacecraft frame polarization, defined relative to the background mean field is directly measurable through the value of $\sigma$ \citep{Narita2009,PodestaGary2011b,Klein2014}. Table \ref{tab:1} provides an overview of the relationship between polarizations measured in the solar wind and spacecraft frames given the Doppler shifted spacecraft frequency
\begin{equation}
    2\pi f =\omega(\bvec{k}) + \bvec{k}\cdot \bvec{V}_{sw}
\end{equation} and assuming that the Taylor hypothesis $\bvec{k}\cdot \bvec{V}_{sw}>>\omega(\bvec{k})$ is applicable.
The importance of this distinction is apparent in the subsequent sections.

\begin{deluxetable}{ccc}
\label{tab:1}
\tabletypesize{\footnotesize}
\caption{Overview of relationship between spacecraft and solar wind frame circular polarization under Taylor's hypothesis for ion resonant wave, e.g. ion cyclotron, and electron resonant, e.g. fast-magnetosonic/whistler, waves.}

\tablecolumns{3}
\tablewidth{0pt}
\tablehead{ Frame & \multicolumn{2}{c} {Polarization ($\sigma$}\\
 & \colhead{Ion Resonant} & \colhead{Electron Resonant}}
 \startdata
 Solar Wind   & + & -\\
 Spacecraft ($\bvec{k}$ $\cdot$ $\bvec{V}_{sw} =1$) & $+$ & $-$\\
 Spacecraft ($\bvec{k}$ $\cdot$ $\bvec{V}_{sw}=-1$) & $-$ & $+$
 \enddata
 \vspace{-0.8cm}
 \end{deluxetable}

\section{Statistics of Wave Events}\label{sec:3}

\begin{figure*}
    \centering
    \includegraphics{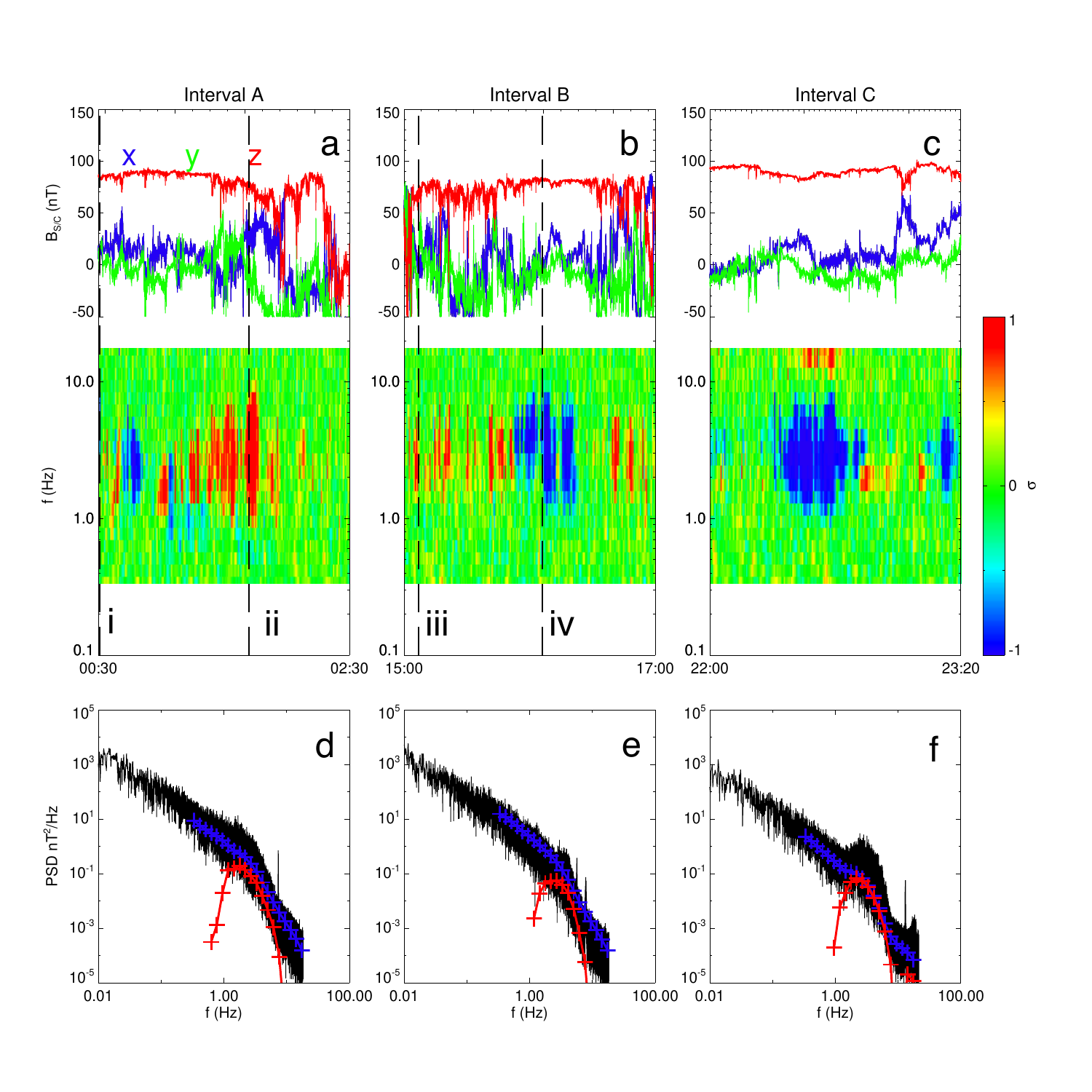}
    \caption{(a-c) Time series for three large coherent wave storms on Nov 05, 2018. Each panel shows $\sigma(f,t)$ the ratio between signed circularly polarized power to the total power of the perpendicular fluctuations (normalized reduced helicity). A value of $\sigma=-1$ (blue) corresponds to an electron resonant rotation relative to the mean field in the spacecraft frame. A  value of $\sigma=1$ (red) corresponds to ion resonant polarization in the spacecraft frame (positive reduced magnetic helicity). (d-f) Trace power spectral densities nT$^2$/Hz of the three intervals. The blue trace shows the corresponding wavelet power spectra $S_0(f)$ computed over each interval. The red trace shows the circular contribution to the power $S_3$. The dashed lines (i-iv) correspond to times of measured distribution functions shown in Figure \ref{fig:10}.}
    \label{fig:2}
\end{figure*}

Figure \ref{fig:2}(a-c) shows several long-duration coherent wave events spanning multiple wavelet scales on Nov 05, 2018. Signatures of mixed spacecraft frame polarization are evident in all three events. Assuming that all waves are of a single wave mode (intrinsic plasma frame polarization), the mixed spacecraft frame polarization at a single frequency indicates the presence of counter-propagating waves. The bottom panels show the power spectrum measured in each interval with the wavelet spectrum overlaid. Figure \ref{fig:2}(d-f) demonstrates that a large fraction of the measured power at ion scales have a signature of circular polarization.

\begin{figure*}
    \centering
    \includegraphics{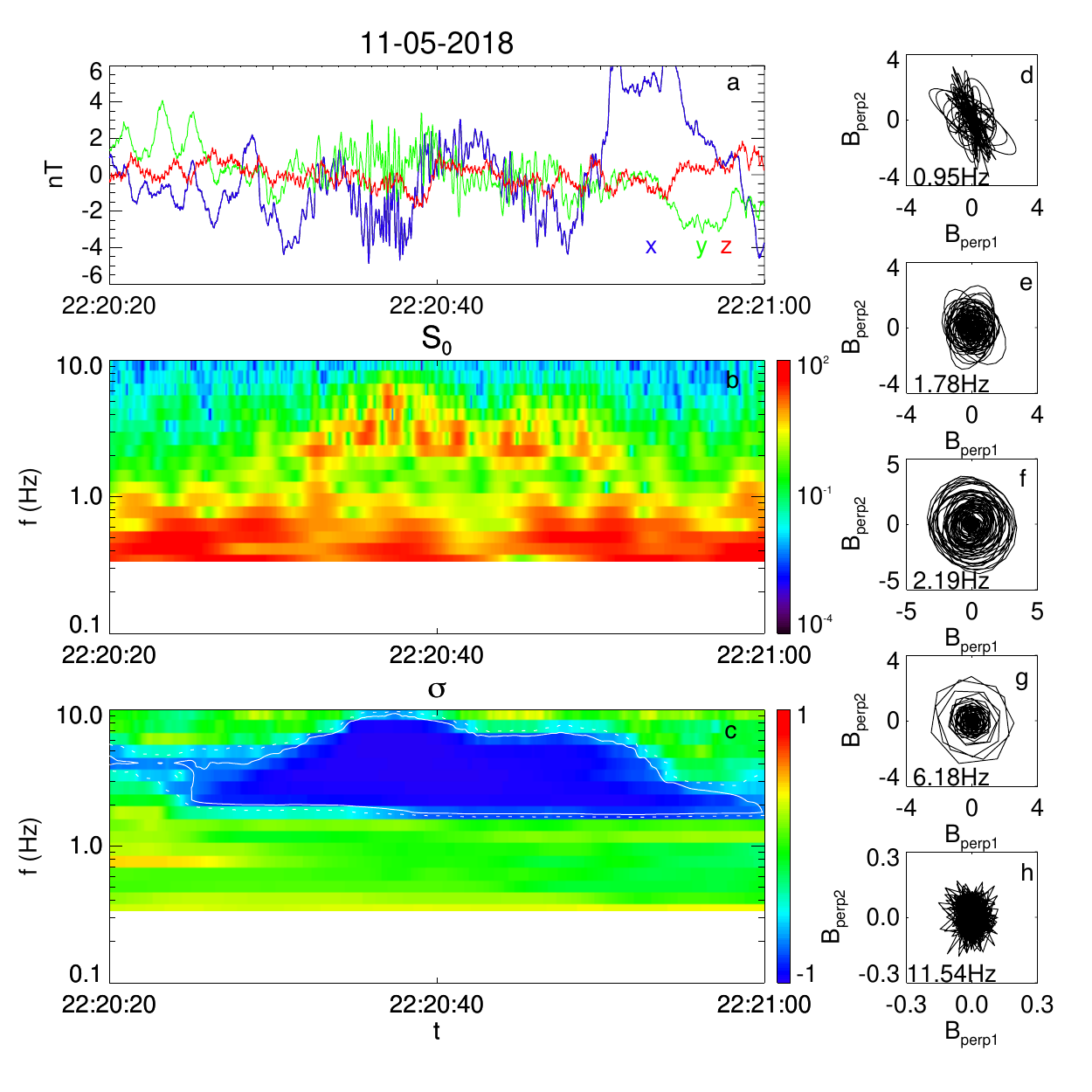}
    \caption{Example of an isolated wave event in background turbulence on Nov 05, 2018 at 22:20:40. (a) Magnetic field time series in spacecraft coordinates (mean subtracted). (b) Wavelet power spectrogram $S_0(t,f)$ for the interval; a broadband event is observed between 2-10 Hz with negative polarization relative to the mean background field in the spacecraft frame. (c) Reduced helicity $\sigma=\langle S_3/S_0\rangle \sim -1$ for the event, indicating an apparent electron resonant polarization. (d-h) Hodograms for $B_{\perp1}$ and $B_{\perp2}$ over a range of scales.}
    \label{fig:3}
\end{figure*}

In addition to large coherent events pictured in Figure \ref{fig:2}, isolated solitary ion scale waves are commonly observed over the encounter. Figure \ref{fig:3}(a) shows an interval on Nov 05, 2018 with coherent fluctuations significantly larger than the background turbulence. A subset of the perpendicular power (i.e.~$S_0$) of the day-long wavelet transform of the interval is presented in Figure \ref{fig:3}(b), and the time-frequency spectrogram of $\sigma$ in Figure \ref{fig:3}(c); a set of hodograms of $B_{\perp1}$ and $B_{\perp2}$ at different wavelet scales is additionally presented. 

In order to characterize the statistics of coherent waves, criteria were developed to separate intervals of coherent power from the turbulent background. At each scale, intervals were identified as a wave event when $|\sigma|>0.7$, the event is extended continuously to neighboring times while $|\sigma|\ge0.5$. The boundaries associated with $|\sigma|>0.7$ and $|\sigma|>0.5$ are shown in Figure \ref{fig:3}(c). While multiple events at different frequencies (wavelet scales) may overlap in time, events at a given scale are necessarily separated from each other. For each event, the total power in the perpendicular fluctuations and duration of the event are measured
\begin{subequations}
\begin{eqnarray}
    S_\sigma=\sum_{i=0}^{N_\sigma-1} S_{0i}\\
    T_\sigma= \sum_{i=0}^{N_\sigma-1} \Delta t_i,
    \end{eqnarray}
\end{subequations}
 where the sum is over the $N_\sigma$ indices between the start and end of each event given by the $|\sigma|>0.5$ threshold.

\begin{figure*}
    \centering
    \includegraphics{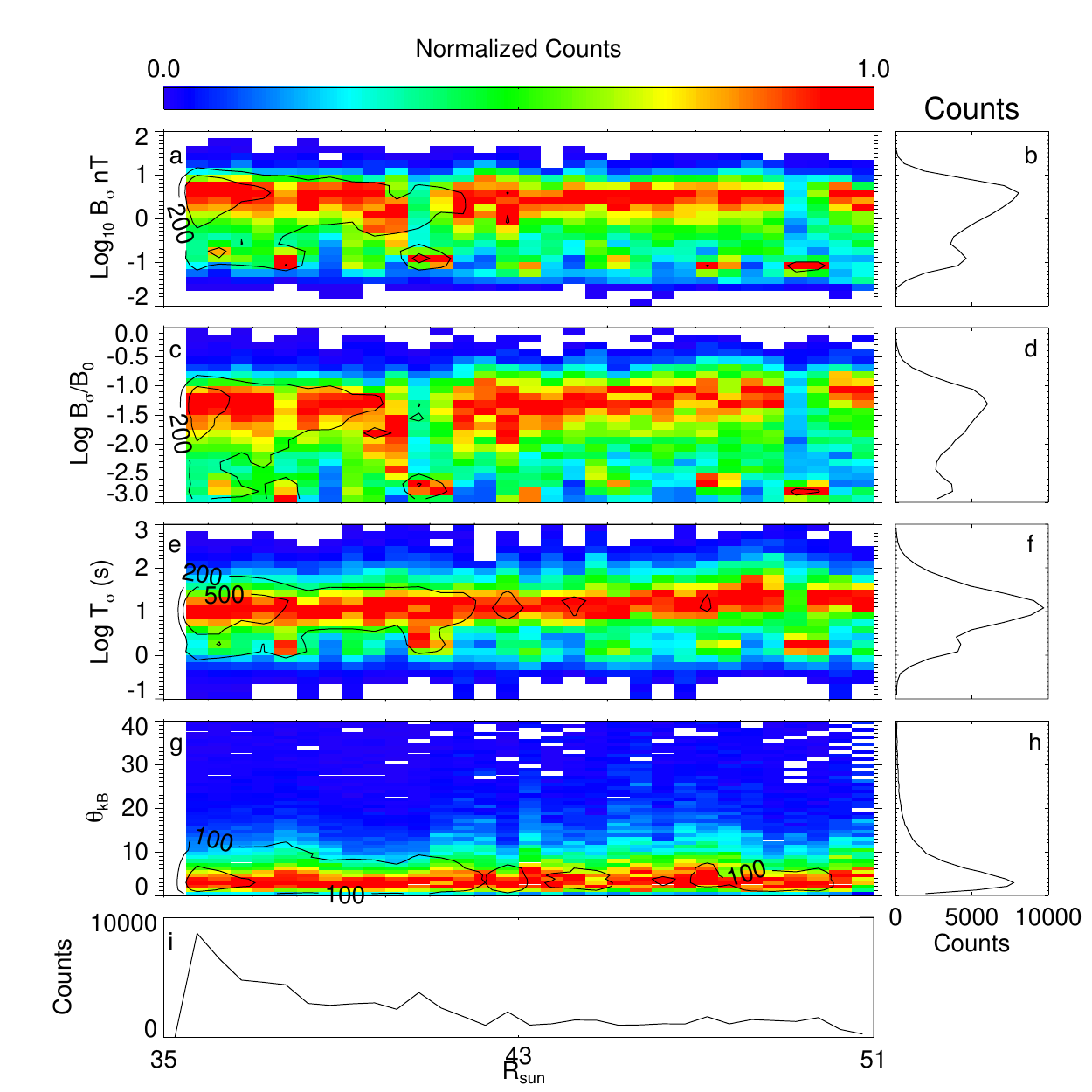}
    \caption{(a) Joint distribution of average wave amplitudes $B_\sigma=\sqrt{S_\sigma/N_\sigma}$ in $nT$ for circularly polarized waves with $R_{\odot}$.  The distribution is normalized to the most probable power at each radius. The average amplitude changes weakly with distance. (b) One dimensional distribution of  $B_\sigma$ corresponding to an rms amplitude 4.22 nT. (c) Joint distribution of the normalized amplitude $B_\sigma/B_0$ with $R_{\odot}$ normalized to the most probable value at each radial radius.(d) One dimensional distribution of $B_\sigma/B_0$ which has a mean of 0.04.  (e) Joint distribution of average wave duration $T_\sigma$ with $R_{\odot}$ normalized to the most probable duration at each radius. (f) One dimensional distribution of $T_\sigma$ which has a mean of 21 s. (g) Joint distribution of coherent wave propagation direction $\theta_{kB}$ in with $R_{\odot}$ normalized to most probable angle at each radius. (h) One dimensional distribution of  $\theta_{kB}$ which has a mean of $5\degr$. (i) One dimensional distribution of counts with $R_{\odot}$, which monotonically decreases. Contours at 200 and 500 count levels are drawn over of the joint distributions (panels a,c,e, and g).}
    \label{fig:4}
\end{figure*}
\subsection{Amplitude \& Duration}\label{sec:3.1}
During E1, 76471 individual events (over all 19 scales) meeting the $|\sigma|>0.7$ threshold were automatically identified. Figure \ref{fig:4}(a) shows the joint distribution of radial distance with average wave amplitude ($B_\sigma$=$\sqrt{S_\sigma/N_\sigma}$), no significant dependence on radius was determined. Figure \ref{fig:4}(b) shows the one dimensional distribution of $B_\sigma$ which is peaked with rms wave amplitude of 4.23 nT. The distribution of fluctuation amplitude normalized to the mean field magnitude, $B_\sigma/B_0$, is shown in Figure \ref{fig:4}(c), with the one dimensional distribution with mean 0.04 shown in Figure \ref{fig:4}(d). Though no dependence is observed in $B_\sigma$ with radius, the scaling of $B_0$ with radius does imply a trend in the normalized $B_\sigma/B_0$. At at 35 $R_{\odot}$ the $B_\sigma/B_0$ distribution is peaked at $B_\sigma/B_0=0.05$, while the distribution at $50 R_{\odot}$ occurs at $B_\sigma/B_0=0.08$. However, the variance of $B_\sigma/B_0$ at any given radius is on order the total radial variation.

Figure \ref{fig:4}(e) shows the joint distribution of the event duration $T_\sigma$ with distance. Figure \ref{fig:4}(f) shows the distribution of event duration with a mean of 22 s. The longest event had a measured duration of 2968.4s and an rms amplitude of 8.2 nT, which occurred with a mean field of 60.9 nT and a $B_\sigma/B_0$ of 0.13.

An estimate of the wave propagation direction is obtained using 
the angle between $B_0$ and the direction associated with the minimum variance direction \citep{Means1972,SonnerupCahill,Santolik2003,Jian2009}.

Figure \ref{fig:4}(g) shows the joint distribution of propagation angle with radius, demonstrating no discernible scaling with distance. Figure \ref{fig:4}(h) shows the distribution of $\theta_{kB}$ between the propagation vector and the mean field with a mean angle of 9.9$\degr$ from the mean field; the median angle is 5.3$\degr$, and the distribution is peaked at $\theta_{kB}=3\degr$. For a pure parallel propagating plane-wave with 4 nT amplitude ( 16nT$^2$ power), fluctuations with rms amplitude < 1 nT are sufficient to introduce a perturbation to the minimum variance direction of this order. Accordingly, we cannot distinguish these waves from perfectly parallel propagating waves.

Figure \ref{fig:4}(i) shows the distribution of event counts as a function of radial distance. While the occurrence of events is much higher at perihelion, the characteristics of the waves do not change drastically with distance. In all panels of Figure \ref{fig:4} the distributions correspond to the total distribution of evets measured at all 19 wavelet scales.

Contrasting the relative lack of scaling of measured wave properties, the background properties of the mean solar wind are measured with strong radial scalings. \cite{Bale2019} show that the mean background magnetic field has the expected $r^{-2}$ scaling. \cite{Huang2019} measure the proton density to scale with $r^{-1.94}$ for the inbound phase of the encounter and a scaling of $r^{-2.44}$ for the outbound. Similarly $T_p$ is measured as $r^{-1.45}$ for the inbound and $r^{-0.90}$ encounter; \cite{Huang2019} . In any case, the radial scalings of background plasma parameters are not evident in the scaling of the coherent wave parameters.

In Figure \ref{fig:4}(b,d) there is a secondary population with an approximate order of magnitude decrease in both amplitude (two orders in power) and duration. Because of the lower occurence rates, as well as low amplitudes and short durations should not severely impact the statistical analysis of the net effect of the dominant ion scale wave signatures. These events may either correspond to kinetic scale turbulence with  non-zero helicity measured at perpendicular angles to the mean field e.g. \cite{Leamon1998,PodestaGary2011,He2011,Klein2014,Woodham2018}, or residual signatures of narrow band reaction wheel noise (e.g. Appendix \ref{appendix}).

\subsection{Wave Frequencies \& Ion Scales}\label{sec:3.2}
For each wave interval, the spacecraft frequencies corresponding to the convected proton gyroscale, $f_\rho$, ion inertial scale, $f_{di}$, and resonant cyclotron scale $f_{res}$ are computed assuming Taylor hypothesis $2\pi f_{sc}=\bf{k}\cdot V_{sw}.$  The proton gyroscale is defined as $$\rho=\frac{m_pV_{\perp pth}}{qB_0},$$ while the inertial scale is related to the ion gyroscale by $d_p=\rho/\sqrt\beta_\perp,$ where $\beta_\perp=V_{\perp pth}^2/V_A^2$ and the Alfv\'en speed is $V_A=B_0/\sqrt{\rho_0\mu_0}$. For protons moving along $B_0$, the cyclotron resonance is given as \begin{equation}
\Omega_p=\omega \pm k_{\parallel} V_{th\parallel} \label{eq:res}
\end{equation} where $\Omega_p=qB_0/m_p$ the $\pm$ relates to the direction of the wave propagation relative to the particles motion \citep{Woodham2018}. 

Using parallel thermal speed from \cite{Huang2019} and the low frequency limit for wave dispersion $\omega \sim k_\parallel V_A$, e.g. \cite{Gary1993book}, gives a resonance condition \begin{equation}k_\Omega=\frac{\Omega_p}{V_A +V_{\parallel pth}}\end{equation}\citep{Leamon1998,Woodham2019}. Using the Taylor hypothesis, spacecraft frequencies corresponding to the Doppler shifted gyroscale, inertial scale and cyclotron resonant scale are given as

\begin{subequations}
\begin{align}
    f_{ci}&=\Omega_p=\frac{qB_0}{m_p}\\
    f_\rho&=\frac{V_{sw}}{2\pi}\frac{qB_0}{m_pV_{\perp pth}}\\
    f_{di}&=\sqrt{\beta}f_\rho\\
    f_\Omega&=\frac{V_{sw}}{2\pi}\frac{\Omega_p}{V_A +V_{\parallel pth}}
    \end{align}
    \label{eq:plasma_scales}
    \end{subequations}

\begin{figure*}
    \centering
    \includegraphics{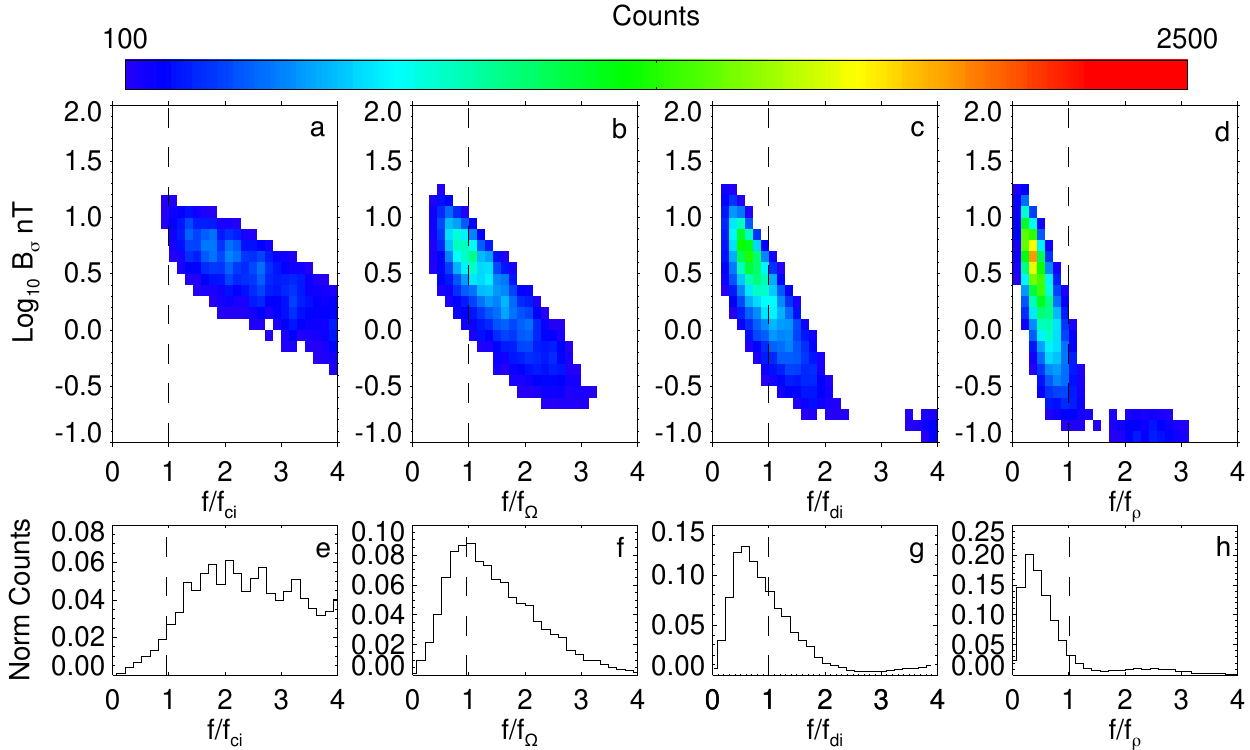}
    \caption{(a-d) Joint distribution of $B_\sigma$ with the wave frequency normalized to spacecraft frequencies corresponding with the ion cyclotron frequency and Doppler shifted cyclotron resonant scale, ion inertial length, and gyroradius: $f/f_ci$, $f/f_{\Omega}$, $f/f_{d}$, $f/f_\rho$. (e-h) Measured distributions of $f/f_{\Omega}$, $f/f_{d}$, $f/f_\rho$. The power is peaked at frequencies corresponding to the cyclotron resonance ($f/f_{res} \sim 1$), and is cutoff at frequencies above the gyroscale ($f/f_{\rho} \sim 1$). Very few events are observed with $f/f_{ci} <1$.}
    \label{fig:5}
\end{figure*}

Figure \ref{fig:5}(a-d) shows the joint distribution of  amplitude $B_\sigma$ with frequency normalized to the ion cyclotron frequency, $\Omega_p$, and each of the ion scales in Equation \ref{eq:plasma_scales}. Figure \ref{fig:5}(e-h) shows the corresponding distributions of event counts at each frequency normalized to the three ion scales. Figure \ref{fig:5}(a) shows that wave power occurs above uniformly above the ion-gyroscale. Figure \ref{fig:5}(b) shows that the wave power is peaked approximately near the Doppler shifted resonant scale $f_{\Omega}$ in the spacecraft frame, while the power is cutoff at the Doppler shifted gyroscale, $f_\rho$ (Figure \ref{fig:5}d). The intermediate ion inertial scale is Figure \ref{fig:5}(c) is shown for completeness. The cutoff at $f_\rho$ is consistent with Vlasov-Maxwell estimates of strong Alfv\'{e}n damping ion-cyclotron at proton kinetic scales \citep{Leamon1998,Gary1999,GaryBorovsky2004}.


Figure \ref{fig:5}(c,d) show that a secondary population of events are present above the Doppler shifted gyroscale, which are consistent with the secondary population from Figure \ref{fig:4}. However, we emphasize these events are separate from the distribution of ion scale waves, and are possibly the result of statistical fluctuations, observations of KAW associated with non-zero helicity measured at perpendicular angles (e.g, \cite{PodestaGary2011,He2011,Woodham2018} or residual signatures of the narrow band reaction wheels (Appendix \ref{appendix}).

\subsection{Angular Dependence of Wave Occurrence and Energy}\label{sec:3.4}
For each scale, the angular dependence of the total energy $S_0(f,\theta_{BV})$ is computed by summing the energy for all times that  $\theta_{BV}$ was within range of angles $\theta_j < \theta_{BV}$.
\begin{equation}
  S_0(f,\theta_{BV})=\sum_{i=0}^{N-1} S_0(f,t_i|\theta_j<\theta_{BV}<\theta_{j+1})
  \end{equation}

The wave contribution to the total observed energy is computed as a function of angle $S^T_{\sigma}(f,\thetabv)$ by summing the energy associated with coherent circularly polarized events, when $\theta_{BV}$ was within range of angles $\theta_j < \theta_{BV} <\theta_{j}+\Delta\theta$. The fraction of energy with positive (proton resonant) and negative (electron resonant) spacecraft frame polarization is further constrained by conditioning on the sign of $\sigma:$
 \begin{subequations}
 \begin{align}
     S^{+}_\sigma(f,\theta_{BV})&=\\
     &\sum_{k=0}^{N-1}S_{\sigma}(f,k|\theta_j<\theta_{BV}<\theta_{j+1}, \text{sgn}(\sigma)=1)\notag\\
       S^{-}_\sigma(f,\theta_{BV})&=\\
     &\sum_{k=0}^{N-1}S_{\sigma}(f,k|\theta_j<\theta_{BV}<\theta_{j+1}, \text{sgn}(\sigma)=-1)\notag\\
     S^T_\sigma(f,\theta_{BV}) &=S^{+}_{\sigma} +{S_{\sigma}^{-}},
 \end{align}
 \end{subequations}
 where the sum $k$ is an index over the set of events at frequency $f$ is conditioned on the angle $\theta_{BV}$ and $|\sigma| >0.7$ as defined in \ref{sec:2}. The notation $\sum f(x|y)$ is understood as the sum of $f$, a function of $x$, conditioned on $y.$ 
 
The angular dependence of the occurrence of the ion scale waves is obtained by comparing the integrated duration of the observed ion scale waves with the angular distribution of $\thetabv.$ The time distribution of $\thetabv$ is measured as the the total time that $\theta_j<\thetabv<\theta_j+\Delta\theta$ regardless of the polarization state:
\begin{equation}
T(f,\theta_{BV})=\sum_{i=0}^{N-1} \Delta t(f|\theta_j<\theta_{BV}<\theta_{j+1})\\
\end{equation}

The wave occurrence rate as a function of scale and angle is determined by integrating the duration of waves, $T_\sigma$, that occur when $\thetabv$ is within the angle bin $\theta_j+\Delta\thetabv$. 
\begin{subequations}
\begin{align}
T^{+}_\sigma(f,\theta_{BV})&=\\
&\sum_{k=0}^{N-1} T_{\sigma}(f,k|\theta_j<\theta_{BV}<\theta_{j+1},\text{sgn}(\sigma)=1)\notag\\
T^{-}_\sigma(f,\theta_{BV})&=\\
&\sum_{k=0}^{N-1} T_{\sigma}(f,k|\theta_j<\theta_{BV}<\theta_{j+1},\text{sgn}(\sigma)=-1)\notag\\
T^T_\sigma(f,\theta_{BV}) &=T^{+}_{\sigma} +{T_{\sigma}^{-}}
\end{align}
 \end{subequations}

The sum over $i$ is taken over the time variable ($t_i=n_i\Delta t$ ) and the sum $k$ over the index of wave events at a given scale.

Figure 7 shows normalized occurrence contours
\begin{equation}
    T^{\pm}(f,\theta_{BV})=T^{\pm}_\sigma(f,\theta_{BV})/T(f,\theta_{BV})
\end{equation}
 for both positive and negative reduced helicity for each full day of perihelion 1 (Nov 01-10, 2018). Contours in red show measured positive (apparent ion-resonant) polarization in the spacecraft frame, while contours in blue show a measured negative (apparent electron-resonant) polarization. The sign of the spacecraft frame polarization is calculated with respect to the mean field direction $B_0(s,t)$ given in Equation \ref{eq:meanb}. Waves propagating outward in the plasma-frame are advected outward by the solar wind, such that measurements in the spacecraft frame are Doppler shifted to higher frequencies and the polarization is maintained in both frames. Conversely, sunward propagating waves are advected outwards by the solar wind, which Doppler shifts spacecraft measurements to lower frequencies. Given the ordering $V_{sw}>V_{A}$ a large Doppler shift through zero frequency will cause the inward propagating plasma frame waves to appear with opposite polarization in the spacecraft frame \citep{Narita2009,HowesQuataert2010,PodestaGary2011b}.

 During the majority of the first encounter PSP is connected to a coronal hole of negative polarity with a large scale magnetic field pointing sunwards such that radial field intervals correspond to an angle of $\thetabv\sim180\degr$ \citep{Bale2019,Badman2019}. During radial field intervals events with apparent positive and negative polarization each occur with  $T_\sigma^\pm \sim10-15\%$ of the time in a range of frequencies from $\sim1-5$Hz. On the inbound phase, there is an equal distribution helical fluctuations with positive and negative polarization such that the total normalized occurrence rate at roughly anti-parallel angles is $T_\sigma^{T}=T_\sigma^++T_\sigma^-~\sim20-30\%$. At perihelion (Nov 06, 2018) and during the outbound portion of the orbit, increased occurrence rates are observed with $T_\sigma^T\sim 75\%$ when there is radial field alignment. On Nov 07, 2018, good statistics are obtained at parallel $\thetabv$, with strong negative reduced helicity evident for parallel field angles. The strong positive measurements at anti-parallel $\thetabv$ and negative measurements at parallel $\thetabv$ suggest that the sign of the wave vector direction may change during the mean field reversal: a similar inversion is observed in the MHD scale cross helicity during radial switchbacks of the mean magnetic field, as a change in direction of the Alfv\'enic flux with respect to a heliocentric coordinate system \citep{DudokdeWit2019,McManus2019}. On Nov 08, 2018 there seems to be enhanced occurrence of wave events with  $T^{+}> 0.75$ at anti-parallel $\thetabv$; such large occurrence rates are not evident in the inbound phase. In all cases there is a cutoff at low frequencies which occurs within the bandwidth of the wavelet transform which is aligned with the value $\Omega_p=qB_0/m_p$. A high frequency cutoff to the events is also observed which increases to higher frequencies at lower altitudes and likely corresponds to the advected gyroscale--e.g. Figure \ref{fig:5}(a,d).

Figure \ref{fig:8} shows normalized power rates $S^{\pm}_3/S_0$
\begin{equation}
    {S_3}^{\pm}(s,\theta_{BV})=S^{\pm}_3(s,\theta_{BV})/S_0(s,\theta_{BV})
    \end{equation}
for each day of E1. 

Though events occur roughly 30-50\% of the time, they contribute a significantly larger fraction to the total power ($S_\sigma^{T}=S_\sigma^+ +S_\sigma^->70\%$), indicating that the integrated contribution to the observed power is much larger than the contribution from background $k_\parallel$ turbulent fluctuations. Intervals with a negative spacecraft frame polarization are commonly observed at higher frequencies than the positive polarization. Intuitively, for counter-propagating ion-resonant waves generated at the same number, sunward propagating are Doppler shifted to negative frequencies in the spacecraft frame, and thus should appear at lower frequencies and opposite helicity than the outward propagating waves, which are Doppler shifted to higher frequencies. Accordingly, if only ICW are present, then the observed distribution of polarizations indicates that different wave-numbers may be excited for counter propagating waves.
 
Observations of negative polarized waves at higher frequencies than positive polarized waves is additionally consistent with electron resonant waves propagating outward. However, it is not clear why the electron resonant waves would be peaked near ion cyclotron resonance, $f_\Omega$, and cutoff at the spacecraft frequency corresponding to the ion gyroscale $f_\rho$.

\begin{figure*}
    \centering
    \includegraphics[width=\textwidth]{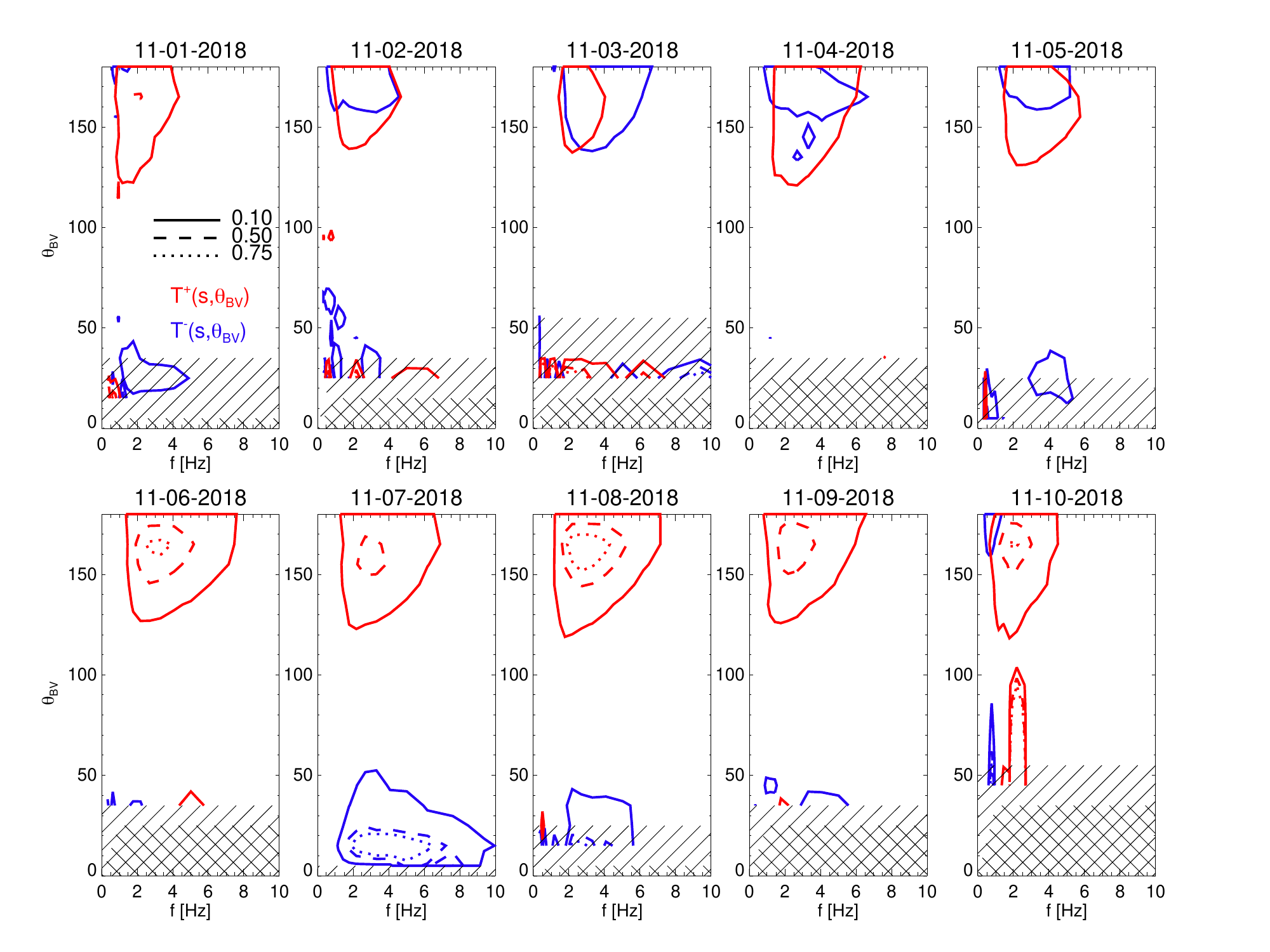}
    \caption{Normalized distribution of occurrence rates of circularly polarized coherent waves over $\thetabv$ and spacecraft frequency for each (full) day of PSP perihelion 1. Three levels of contours correspond to presence of waves in 10, 50 75\% of intervals with the angle between mean magnetic field and solar wind flow $\theta_j<\thetabv<\theta_{j+1}$, where $\theta_{j+1}-\theta_{j}= 10\degr$. Red contours show occurrence of ion resonant waves (positive helicity in spacecraft frame), and blue contours show occurrence rates of waves with apparent electron resonance (negative helicity in spacecraft frame). Sets of slanted hashed are used to identify angle bins accounting for  $<1\%$ of the data for each day, and when no measurements in angle bin were made.}
    \label{fig:7}
\end{figure*}



\section{Observability of Parallel Propagating Waves in an Anisotropic Turbulent Cascade}\label{sec:4}

Figures \ref{fig:7}-\ref{fig:8} reveal a statistical preference for the occurrence of coherent waves during radial field intervals, a result consistent with observations of coherent waves at larger heliospheric distances \citep{Murphy1995,Jian2009,Jian2014,Boardsen2015}. However, this preference for radial field is possibly due to measurement effects. Observational evidence demonstrates that the anisotropy of solar wind turbulence leads to larger amplitude fluctuations perpendicular to the mean magnetic field \citep{Chen2010,Horbury2012}. The increased turbulent power at perpendicular angles may obscure coherent wave power. Additionally, constraints imposed by single point spacecraft measurements preclude a full vector measurement of the spectral density $P(\bvec{k})$,  such that $k_\parallel$ structures may not be observable at oblique $\thetabv$. 

To test these effects we consider the reduced spectrum measured by a single spacecraft:

\begin{equation}
E(f,\theta_{VB})=\int d^3\bvec{k} P(\bvec{k})\delta[2\pi f -(\bvec{k}\cdot{\bvec{V}_{sw}} +\omega)] \label{eq:reduced_spect}
\end{equation}

\cite{FredricksCoroniti1976}.

The Taylor hypothesis corresponds to the limit $\bvec{k}\cdot\bvec{V}_{sw} >>\omega$. Similarly the sign of the measured helical fluctuations corresponds to the reduced magnetic helicity measured along the sampling direction of the Doppler shifted fluctuations in the spacecraft frame \citep{MatthaeusGoldstein1982,Narita2009,HowesQuataert2010}.
 Accordingly, observation of parallel propagating waves may be inhibited when the turbulent background fluctuations are sufficiently large (which is more likely to occur when sampling perpendicular cascade), or when the angle between the sampling direction and the mean magnetic field is sufficiently oblique such that polarization plane of the transverse waves is out of the plane defined by the reduced helicity. 

A simple model spectrum is constructed to test the observability of parallel propagating waves at various angles of $\theta_{BV}$ in the presence of an anisotropic turbulent background. The turbulence is modeled as a slab of parallel propagating ICW waves with a 2D background of axial symmetric perpendicular turbulence 
\begin{equation}
    P(\bvec{k})=P_{2D}(k_\perp) +P_{ICW}(k_\parallel)
\end{equation}
\citep{Bieber1996}. While in principle the contribution of either slab turbulence with $k_\perp=0, k_\parallel \neq 0$, or a critically balanced turbulent spectrum $k_\perp >> k_\parallel$ may be included, the observed dominance of the coherent wave power (Figure \ref{fig:8}) suggests that this simplified model is sufficient. 

The perpendicular turbulent spectrum is modeled as

\begin{equation}
P_{2D}(k_\perp)= A_{2D}k_\perp^{-(1+ \alpha)} \delta(k_\parallel)
\end{equation}
 which, using Equation \ref{eq:reduced_spect}, gives a reduced spectrum 
 \begin{equation}
 E_{2D}(f,\theta_{VB})= C_{2D}f^{-\alpha} \text{sin}\theta_{VB}^{\alpha-1}\label{eq:red2dturb}
 \end{equation}
where $A_{2D}$ and $C_{2D}$ are constants and the Taylor hypothesis is assumed \citep{Bieber1996,Horbury2008,Forman2011}.

Observations of the turbulent spectrum at $\theta_{VB} \approx 90\degr$ constrains both the spectral index $\alpha$ and the coeffecient $C_{2D}$ such that Equation \ref{eq:red2dturb} predicts the reduced turbulent spectrum at various $\theta_{BV}$ as

\begin{equation}
E^*_{2D}(f,\theta_{VB})=  E(f,90{\degr})\text{sin}\thetabv^{\alpha-1}. \label{eq:2dturb}
\end{equation}
Figure \ref{fig:9} a shows the observed wavelet power spectra $S_0(f,\theta_{BV})$ on Nov 05 2018. As expected, the perpendicular spectrum, $S_0(f,80\degr<\theta_{BV}<90\degr)$ demonstrates the largest power consistent with $k_\perp >> k_\parallel$ anisotropy. Because of axial symmetry the angular dependence of the power spectra is restricted to $0<\theta_{BV}<90{\degr}$ and the supplementary angle of $\theta_{BV}$ is used when $\theta_{BV}> 90$ in order to improve statistics. 

Figure \ref{fig:9}(b) shows synthetic turbulent power spectra $E^*_{2D} (f,\theta_{BV})$ for $\thetabv <90\degr$ using measurements of $S_0(f,80\degr<\theta_{BV}<90\degr)$ and Equation \ref{eq:2dturb}, with the empirically measured $\alpha=-1.9$.

The effect of single point measurement effects on observing narrow-band coherent waves with dominant parallel wave-numbers is determined by evaluating the reduced spectra corresponding to spacecraft frame observations of a $\delta$ function spectrum:

\begin{subequations}
\begin{align}
P_{\delta\parallel}({\bf{k}})&=A_\parallel[\delta(k_{0\parallel}-k_\parallel)\delta(k_\perp)]\\
E_{\delta\parallel}(f,\theta_{BV})&=\frac{A_\parallel}{V_{sw}\text{cos}{\theta_{BV}}}\delta\left(k_{0\parallel} -\frac{2\pi f}{V_{sw}\text{cos}\theta_{BV}}\right). \label{eq:deltapar}
\end{align}
\end{subequations}
Equation \ref{eq:deltapar} gives the reduced energy spectrum of a single wave mode wave-vector {\bf{k}}=$k_{0\parallel}
\hat{B_0}$ as a function of spacecraft frequency and $\theta_{BV}$, Figures \ref{fig:7}-\ref{fig:8} show that a broadband spectrum of parallel waves is typically observed. An estimate of the parallel spectrum of waves $P_{\parallel}(k_\parallel)$ is empirically constructed from observations of the contribution circularly polarized power to the wavelet spectrum when $0<\theta_{VB}<10^{\degr}$ on Nov 05, 2018. Using $k_{0\parallel}\approx 2\pi f/V_{sw}$ the parallel spectrum is modeled as a superposition of weighted $\delta$-functions at each wavelet scale \begin{equation}
    P_{\parallel}(k_\parallel)=\sum S^T_\sigma(f,0)\delta(2\pi f/V_{sw} -k_\parallel).
    \end{equation}

A synthetic reduced spectrum at oblique $\theta_{VB}$ for the parallel ICW events  $E^*_{\parallel}(f,\theta)$ observed at oblique angles is computed using Equation \ref{eq:deltapar} and the set of weighted $\delta$ functions to fit the parallel spectrum at $\thetabv=0$.
Figure \ref{fig:9}(c) shows the measured angular distribution of circularly polarized power at each frequency $S^T_\sigma(f,\theta_{BV})$. Figure \ref{fig:9}(d) shows the synthetic reduced spectrum of circularly polarized power  $E^*_{\parallel}(f,\theta)$ using   $S^T_\sigma(f,0)$ and Equation \ref{eq:deltapar}. 

Figure \ref{fig:9}(e) shows the fractional polarized power of $S^T_{\sigma}(f,\thetabv)/S_0(f,\thetabv)$ measured on Nov 05, 2018. The corresponding ratio of synthetic reduced spectra $E^*_{\parallel}/E^*_{2D}$ is shown in Figure \ref{fig:9}(f). Qualitatively similar evolution of the distribution of circularly polarized power is observed in both the observations and the synthetic reduced spectra, demonstrating that the disappearance of circular polarization signatures at oblique angles is consistent with sampling effect due to single point measurements of a quasi-parallel wavevector at oblique angles as well as the increased amplitudes of the anisotropic turbulence.

\section{Comparison with Plasma Properties}\label{sec:5}

Several authors have suggested that ion scale waves observed at 1 AU are driven by kinetic instabilities due to temperature anisotropies and beaming secondary proton and $\alpha$ particle populations \citep{Jian2009,PodestaGary2011b,Klein2014,Wicks2016,Klein2018,Zhao2019,Woodham2019}. 
The SPC Faraday cup instrument on PSP measures a 1-D (reduced) velocity distribution function in the sun-pointing direction. Figure \ref{fig:10}(a-d) shows four distribution functions from SPC corresponding to intervals in Figure \ref{fig:2}; two Maxwellian fits are performed corresponding to drifting proton distributions. Additionally, fits to $\alpha$-particle distribution are shown in units of proton-equivalent speed (i.e. $\sqrt{2}$ times the proton speed due to the charge-to-mass ratio). Figure \ref{fig:10}(a) shows the VDF at the beginning of the interval in Figure \ref{fig:2}(a) where the fraction of circular power is low, while Figure \ref{fig:10}(b) corresponds to large circularly polarized component the cyclotron storm shown in Figure \ref{fig:2}(a); the drift between two proton populations is observed in Figure \ref{fig:10}(b) may contribute to the growth of coherent waves. 

However, Figure \ref{fig:10}(c) shows the measured VDF during an interval of low circular power taken from Interval B in Figure \ref{fig:2}(b). Though some relative drift is observed between the populations, little polarized power is observed. Figure \ref{fig:10}(d) shows the VDF of an interval with large circular polarization from Interval B in Figure \ref{fig:2}(b) with smaller relative drift between proton populations, suggesting that the proton beam drift may not drive the distribution unstable in this event. However, there is a slight shift in the peak of the $\alpha$-particle distribution, possibly indicating that streaming $\alpha$-particles may contribute to these waves. Additionally, it is important to note that a single SPC measurement returns a reduced one dimensional distribution and does not recover the full 3-D distribution of the plasma, which may reveal large temperature anisotropies $T_\perp/T_\parallel$  in either the core or beam population commonly associated with electromagnetic ion cyclotron instabilities \citep{Gary1993book,PodestaGary2011b}.

By integrating temperature observations from SWEAP/SPC and magnetic fields from FIELDS, \cite{Huang2019} estimate the proton temperature anisotropies in one minute sampling intervals with 10 second cadence. The the normalized polarization $\sigma(f)$ is computed for each 10 second integration for each wavelet scale. The largest positive value of $\sigma(f)^+$ is taken as a measure of ion-resonant waves, while the largest negative value $\sigma(f)^-$  is measure of an electron-resonant polarization.  Figure\ref{fig:11}(a-b) shows joint probability distributions in the $\beta_\parallel-T_\perp/T_\parallel$ plane using the \cite{Huang2019} data set colored by $\sigma^{\pm}$. Contours in either plot show the distribution of measured wave events of the corresponding polarization. Additionally, instability thresholds for the Alfv\'{e}n ion-cyclotron instability, $T_\perp/T_\parallel>1$, and parallel fire-hose, which drives fast magnetosonic/whistlers at $T_\perp/T_\parallel<1$, are drawn at $\gamma/\Omega_p =10^{-2}$ and $\gamma/\Omega_p =10^{-4}$ using fit parameters determined in \cite{Verscharen2016}. A statistical preference for ion-resonant polarization appears at $T_\perp/T_\parallel>1$, and a secondary population of electron-resonant polarization appears at $T_\perp/T_\parallel<1$ consistent with the results of \cite{Woodham2019} and \cite{Zhao2019} at 1 AU. However, the distribution of measured ion-scale waves is not particularly bound into any region of the $\beta_\parallel-T_\perp/T_\parallel$ plane. Additionally the presence of electron-resonant waves occurs at significantly lower values of $\beta$ than what would be suggested by the fast magnetosonic/whistler waves driven by the fire-hose instability derived in \cite{Verscharen2016}. Further work is required to determine the effect of the secondary proton and $\alpha$ populations.

\section{Independence from Electron Scale Waves}\label{sec:6}

\cite{Malaspina2019} demonstrate that electrostatic waves near electron cyclotron scales made by PSP/FIELDS occur preferentially during intervals with radial magnetic field. The occurrence of electron waves is parameterized by the electron wave counts per minute $n_{elec}$. In order to compare the occurrence of ion scale wave events with high-frequency electron waves, the fractional circular power, $\bar{S_3}$, is computed on the same minute time base as the \cite{Malaspina2019} electron counts.

The probability of an ion scale wave $P_{ion}=P(\bar{S}_3>>0.05)$ is computed over the 10 day interval on the minute-cadence time base. Additionally the probability of an electron wave count $P_{elec}=P(n_{elec}>1)$ is determined. Individually, the probability of an ion wave event is $P_{ion}\sim 0.17$ and of an electron event is $P_{elec} \sim 0.15$. The joint probability of observing both an ion scale and electron scale event is $P(n_{elec} >1,\bar{S}_3>0.05) \approx 0.04$, which is approximately equal to the value of $P_{ion}P_{elec} =0.03$, indicating a lack of correlation between the events. Additionally, when conditioning on radial intervals such that $\thetabv <25\degr$ the  probabilities are 

$$P^{\parallel}_{ion}=P(\bar{S}_3>0.05 |\thetabv <25\degr)=0.38$$
$$P^{\parallel}_{elec}=P(n_{elec}>1) |\thetabv <25\degr)=0.35.$$ 

However the joint probability of both an electron wave count and a large ion scale polarization is $$P(n_{elec} >1,\bar{S}_3>0.05)|\thetabv <25\degr) = 0.12$$ which is approximately that of the probability $$P^{\parallel}_{ion}P^{\parallel}_{elec}=0.13$$ indicating uncorrelated distributions.

While both ion scale and electron scale waves have similar probabilities of occurrence, the joint distribution suggests that the occurrence of events is uncorrelated. Perhaps this result is not particularly surprising as the ion instabilities which generate low frequency waves at several Hz are likely decoupled from the electron instabilities acting at kHz, the independent occurrence of these waves reinforces that the young solar wind is subject to, and capable of maintaining, multiple instabilities acting on different scales which have {\em{in-situ}} electromagnetic signatures.

\begin{figure*}
    \centering
    \includegraphics[width=\textwidth]{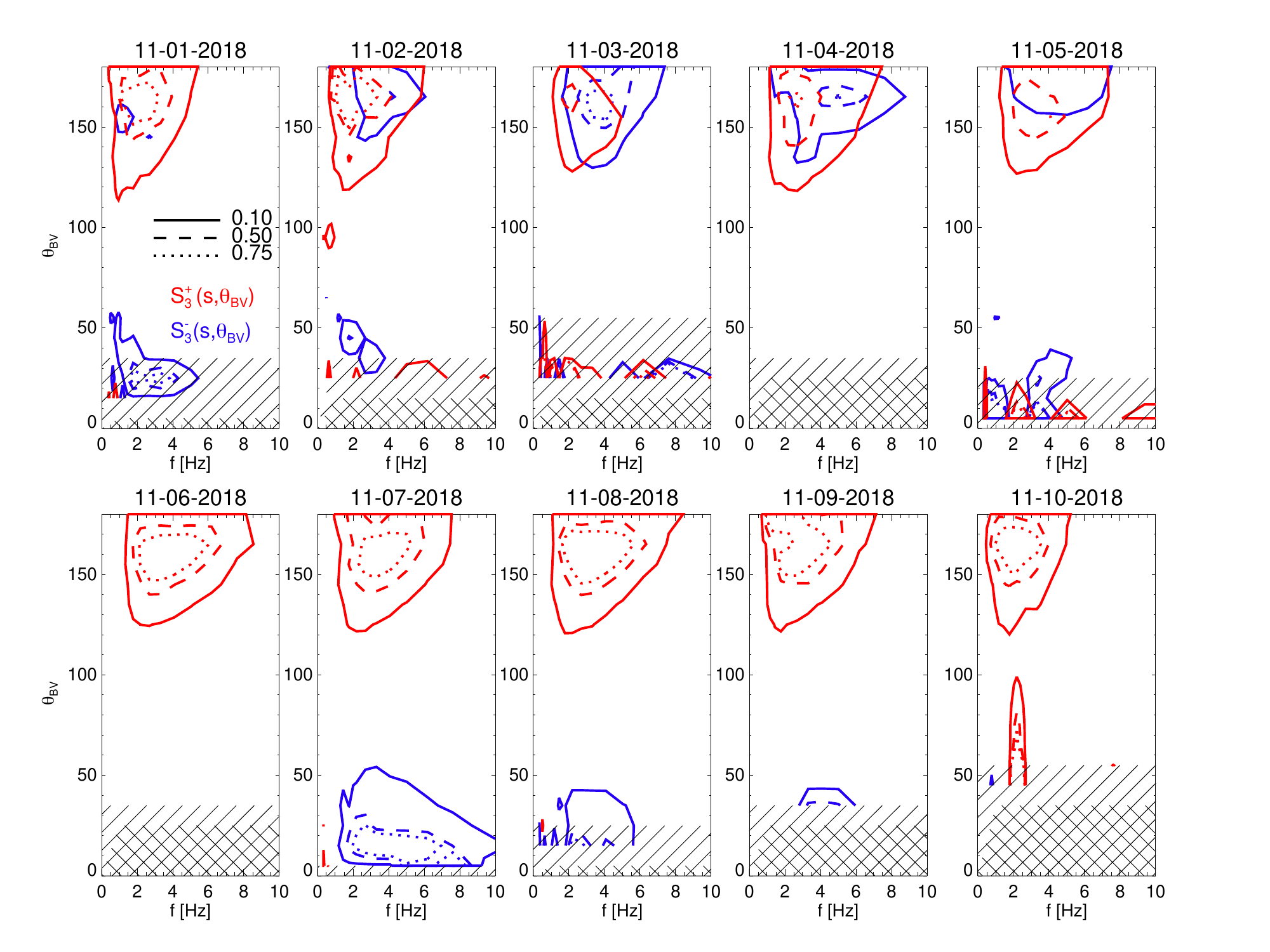}
    \caption{Distribution of energy in circularly polarized coherent waves relative to total observed energy as function of $\thetabv$ and spacecraft frequency for each (full) day of PSP perihelion 1. Three levels of contours correspond to 10, 50 75\% levels of fractional wave power relative to total measured energy when $\theta_j<\thetabv<\theta_{j+1}$, where $\theta_{j+1}-\theta_{j}= 10\degr$. Red contours show occurrence of polarization with apparent ion resonance waves (positive helicity in spacecraft frame), and blue contours show occurrence rates of waves with apparent electron resonance (negative helicity in spacecraft frame). Sets of slanted hashed are used to identify angle bins accounting for  $<1\%$ of the data for each day, and when no measurements in angle bin were made.}
    \label{fig:8}
\end{figure*}

\begin{figure*}
\centering
   \includegraphics{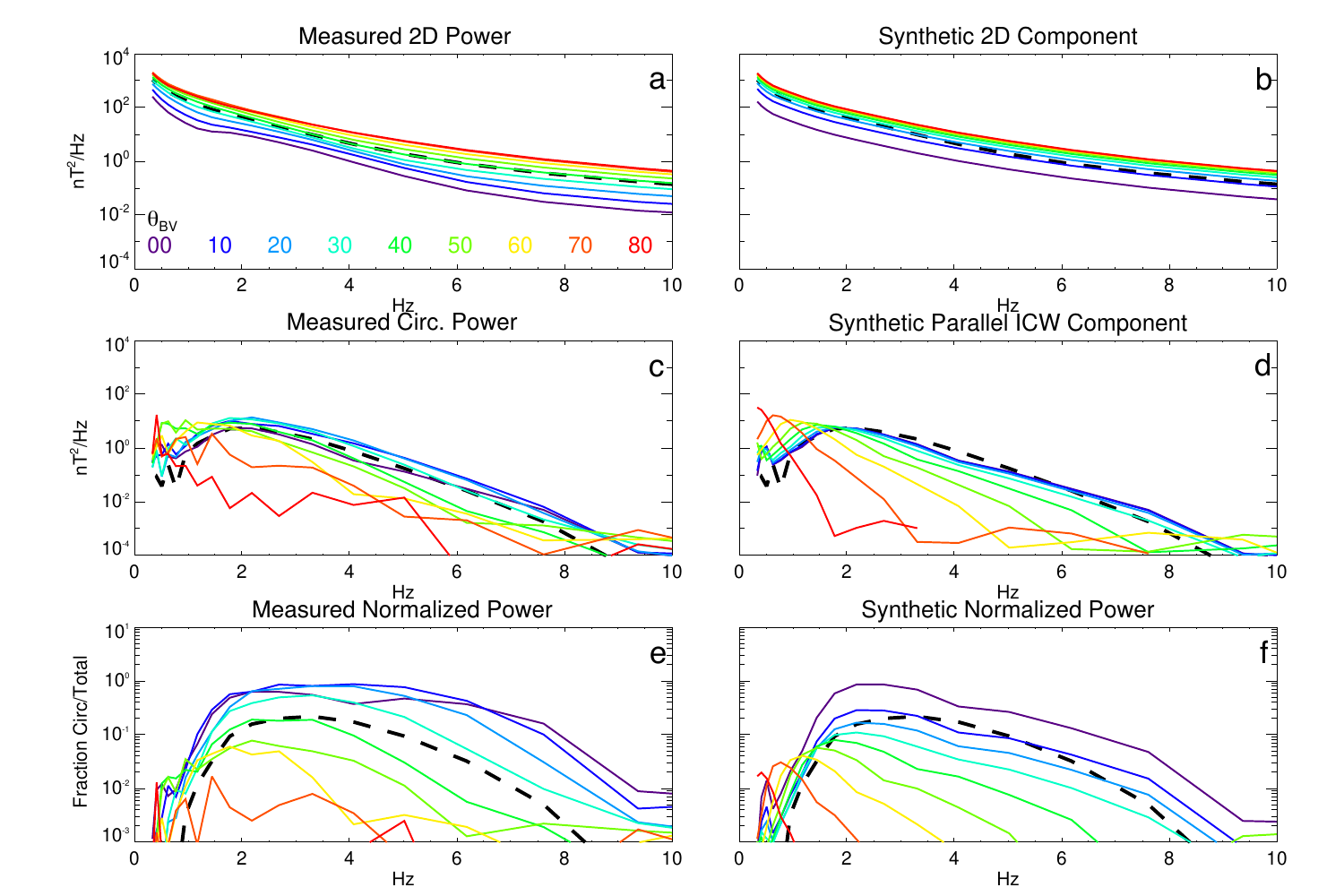}
   \caption{(a) Observations of reduced spectra on Nov 05, 2018 at various angles of $\thetabv$. (b) Synthetic two dimensional component of background turbulent fluctuations assuming $k_\perp$ spectra at oblique angles using observations of reduced spectra at $\thetabv=90\degr$. (c) Observations of reduced spectra at various $\thetabv$ with large circular polarization. (d) Synthesized component of circular polarized power at oblique $\thetabv$ using observations at $\thetabv=0\degr$ and assumption of $k_\parallel$. (e) Fraction of circular power at various angles of $\thetabv$. (f) Fraction of circularly polarized power at various $\thetabv$ using ratio of synthetic parallel and perpendicular spectra.}
    \label{fig:9}
\end{figure*}

\section{Discussion}

Coherent ion scale waves in the heliosphere, which dominate $k_\parallel$ fluctuations, are likely a signature of processes connected to solar wind heating and acceleration through linear resonance and non-linear instabilities \citep{Leamon1998,GaryBorovsky2004,HollwegIsenberg2002,Wicks2016,Zhao2019,Woodham2019}. Previously, \cite{PodestaGary2011} and \cite{He2011} identified signatures of magnetic helicity at parallel $\thetabv$ near ion scales without simultaneous identification of a separate (non-power law) component in the observed spectra, suggesting the presence of low-level background quasi-parallel waves which interact with the turbulent cascade. In contrast, large amplitude coherent events have been thoroughly studied with substantial evidence for generation through plasma instabilities \citep{PodestaGary2011b, Jian2014, Wicks2016, Zhao2019}.

While large amplitude coherent wave events are observed at 1 AU (and elsewhere in the heliosphere), their occurrence is significantly enhanced in the inner heliosphere \citep{Jian2009,Jian2010,Boardsen2015, Zhao2018}. The statistical analysis of polarization signatures in Section \ref{sec:3.1} suggests that typical ion scale waves are quasi-parallel structures with rms amplitudes of $\sim$ 4 nT, which last on order 10 seconds to 1 minute; though the longest coherent wave events may exist on order an hour. 

Section \ref{sec:3.2} shows that the observed waves are well confined to spacecraft frame frequencies between the ion cyclotron frequency $f_{ci}=qB_0/m$ and Doppler shifted proton-gyroscale, $f_{\rho}=\rho V_{sw}/2\pi$. Additionally, the measured distribution of waves is peaked at the spacecraft frequency corresponding to resonant ion-cylcotron interactions \citep{Leamon1998,Woodham2018}. The localization of waves to ion scales suggests that a large fraction of the waves are ion resonant. 

Section \ref{sec:3.4} demonstrates that coherent waves occur in 30-50\% of intervals with approximately radial mean magnetic field configurations. However, the analysis in Section \ref{sec:4} shows that the observability of these events is strongly dependent on the amplitude of the background turbulence and the angle $\thetabv$. The preferential occurrence of waves with a radial field alignment, though consistent with observations from the outer heliosphere, e.g. \cite{Murphy1995, Jian2009,Boardsen2015}, is consistent with sampling effects related to single point measurements of parallel wave number structures in a quasi-perpendicular turbulent cascade. Accordingly, we cannot exclude the possibility that ion scale waves are present during intervals with a non-radial magnetic field configuration. This result suggests that coherent ion scale waves at 1 AU are possibly more common than currently thought, as intervals of radial field are less frequently encountered at 1 AU due to the mean magnetic field direction along the Parker spiral.

Analysis of plasma distribution functions measured by SPC in Section \ref{sec:5} suggests that temperature anisotropy plays a role in the generation of ion scale waves. Using measurements of the proton core temperature anisotropy by \cite{Huang2019}, we find that ion resonant fluctuations occur predominantly when the core proton temperature anisotropy $T_\perp/T_\parallel >1 $ and electron resonant fluctuations occur with $T_\perp/T_\parallel <1$. This result is consistent with observations at 1 AU by \cite{Woodham2019,Zhao2019}, and suggestive of generation of ICW events through Alfven/ion cyclotron instability. However, the observed distribution of  wave events of either polarization does not seem bounded by any portion of the $\beta-T_\perp/T_\parallel$ parameter space. This suggests that additional sources of free energy--e.g. beams and drifts--may be responsible for the growth and generation of these waves. A full analysis of beam and $\alpha$-particle drifts is required to understand the generation of these events through instabilities. Our future work will incorporate the statistics of ion-scale waves with analysis of the full multi-population 3D plasma distribution.



The events measured by PSP tend to have amplitudes $\sim$ nT  and do not scale strongly with radius. Due to the observed scaling in radial turbulent properties reported in \cite{Chen2019}, these results suggest that the observed ion scale waves are more in common with instability driven events rather than an ambient quasi-parallel population of waves interacting with the background turbulence. However, the presence of an ambient population of cyclotron waves associated with the ion-cyclotron damping of the Alfv\'{e}nic turbulent cascade is not ruled out by this study \citep{Leamon1998,Woodham2018}. Additionally, the lack of strong radial scaling suggests that the events are not signatures of near-sun cyclotron heating and are generated by {\em{in-situ}} processes. This conjecture is additionally supported by the analysis in Section \ref{sec:3.2} which shows that wave power is peaked at local values of the cyclotron resonance, and cutoff above local proton gyroscale, where strong cyclotron damping is expected \citep{Leamon1998,Gary1999}. The weak dependence of duration and amplitude with distance may suggest that the processes which generate the individual wave events do not vary greatly over $35-50 R_{\odot}$. 


Observations of circularly polarized magnetic fluctuations in the spacecraft frame can not uniquely determine the polarization in the solar-wind frame \citep{Narita2009,HowesQuataert2010}. Coherent waves with an observed ion-resonant polarization in the spacecraft frame may be associated with either intrinsically ion resonant waves propagating outward or inward propagating electron waves which are Doppler shifted in the spacecraft frame \citep{He2011,PodestaGary2011b,Klein2014,RobertsLi2015,Woodham2019}. Conversely,
observations of electron polarized events correspond to outward propagating electron-resonant modes or the Dopper shift of inward propagating ion-resonant modes.

The Doppler shift of counter propagating waves generated at the same plasma-frame wave-number causes a frequency splitting of the waves in the spacecraft frame, with the inward propagating waves occurring at lower frequencies. For a population of purely ion-resonant cyclotron waves, the inward propagating waves appear at lower frequency and with an electron-resonant polarization. However, waves with electron resonant helicity commonly appear at higher frequencies than the proton resonant helicity, e.g. Figure \ref{fig:8} Nov 3-4, 2018. This suggests that both ion cyclotron and electron fast-magnetosonic/whistler waves may be present in these observations. Additionally, Figure \ref{fig:10} shows that there is a statistical preference for electron polarization with a $T_\parallel/T_\perp <1 $ and ion polarization for $T_\perp/T_\parallel >1$. Future work to compare the observed polarization of electric field fluctuations simultaneously with magnetic fluctuations will provide a definitive measurement of the plasma-frame polarization of the observed waves \citep{Santolik2003}.  

Though this work focuses on the dynamics of the protons, $\alpha$ particles and heavy ions can play a significant role shaping the dispersion of ion-scale waves \citep{HollwegIsenberg2002}. Observations of collisional processing of $\alpha$ particles at 1 AU suggests that preferential ion heating may exist out to $20R_\odot$ to $40R_\odot$, a range now explored by the PSP mission \citep{Kasper2017}. Understanding the effects of the full 3D drifting distribution function of protons and minor ions is imperative in determining the role of these waves in solar wind heating and acceleration.

\section{Summary}

Our results demonstrate the ubiquitous presence of ion-scale waves in the young solar wind. The waves are commonly found at scales coincident with proton-cyclotron resonance, and are cutoff at the proton-gyroscale. A weak radial scaling of the events is observed, indicating that events are likely generated through {\em{in-situ}} processes. Analysis of core proton distribution functions suggests that temperature anisotropy may drive the waves, though we have not yet considered full distributions with drifting secondary proton and $\alpha$  populations. 


Ion scale waves are preferentially observed during alignment between the mean magnetic field and solar wind flow direction, consistent with observations further out in the heliosphere \citep{Murphy1995,Jian2009,Jian2010,Jian2014}. However an analysis of the reduced spectra associated with quasi-parallel coherent structures made from single point spacecraft measurements in an anisotropic turbulent background reveals that the disappearance of wave events at oblique angles is consistent with measurement effects. Accordingly, it is likely that coherent ion-scale waves driven by instabilities are present in the solar wind even during non-radial field intervals.

\begin{figure*}
    \centering
    \begin{minipage}{\columnwidth}
        \includegraphics[width=\columnwidth]{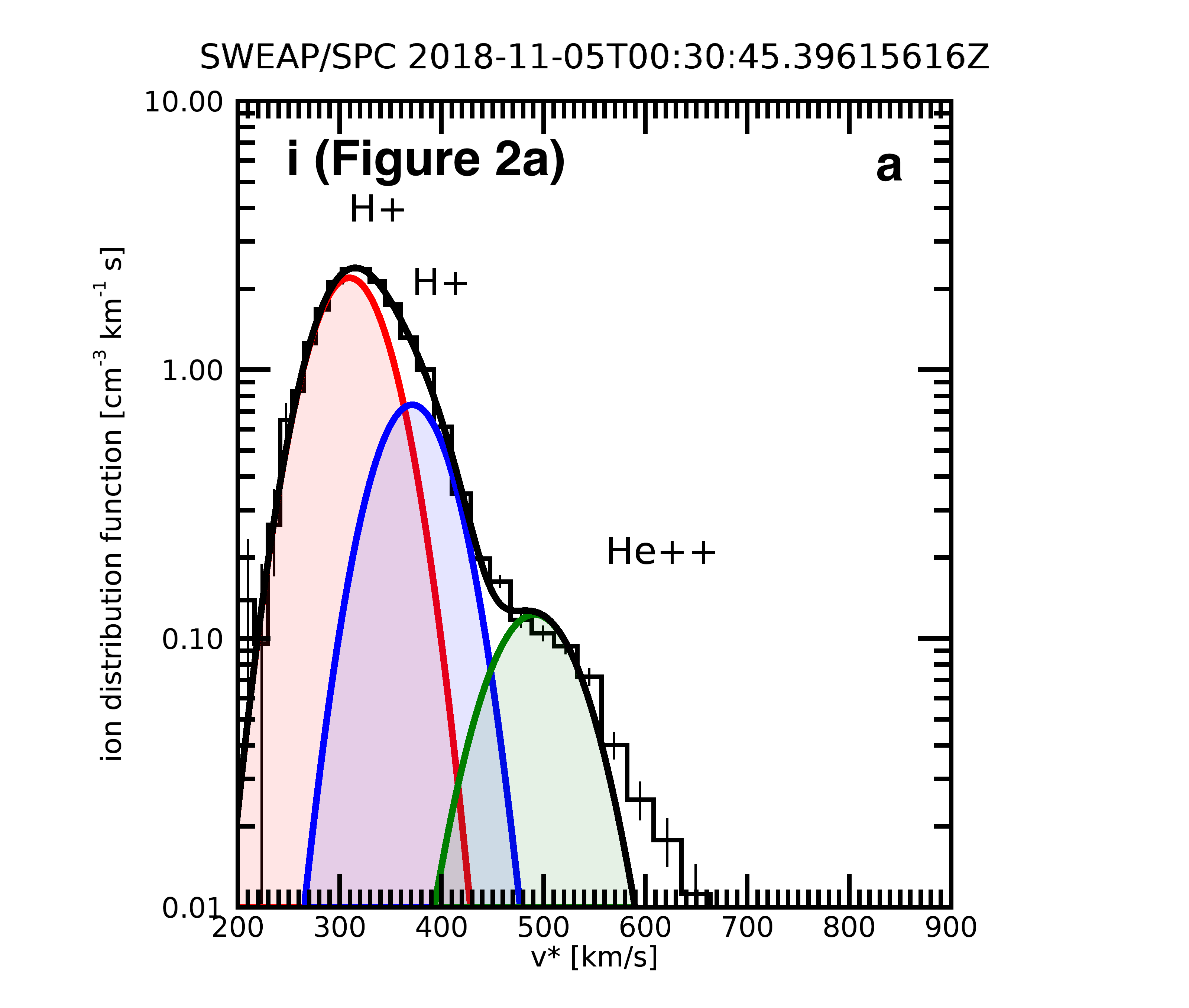}
        \includegraphics[width=\columnwidth]{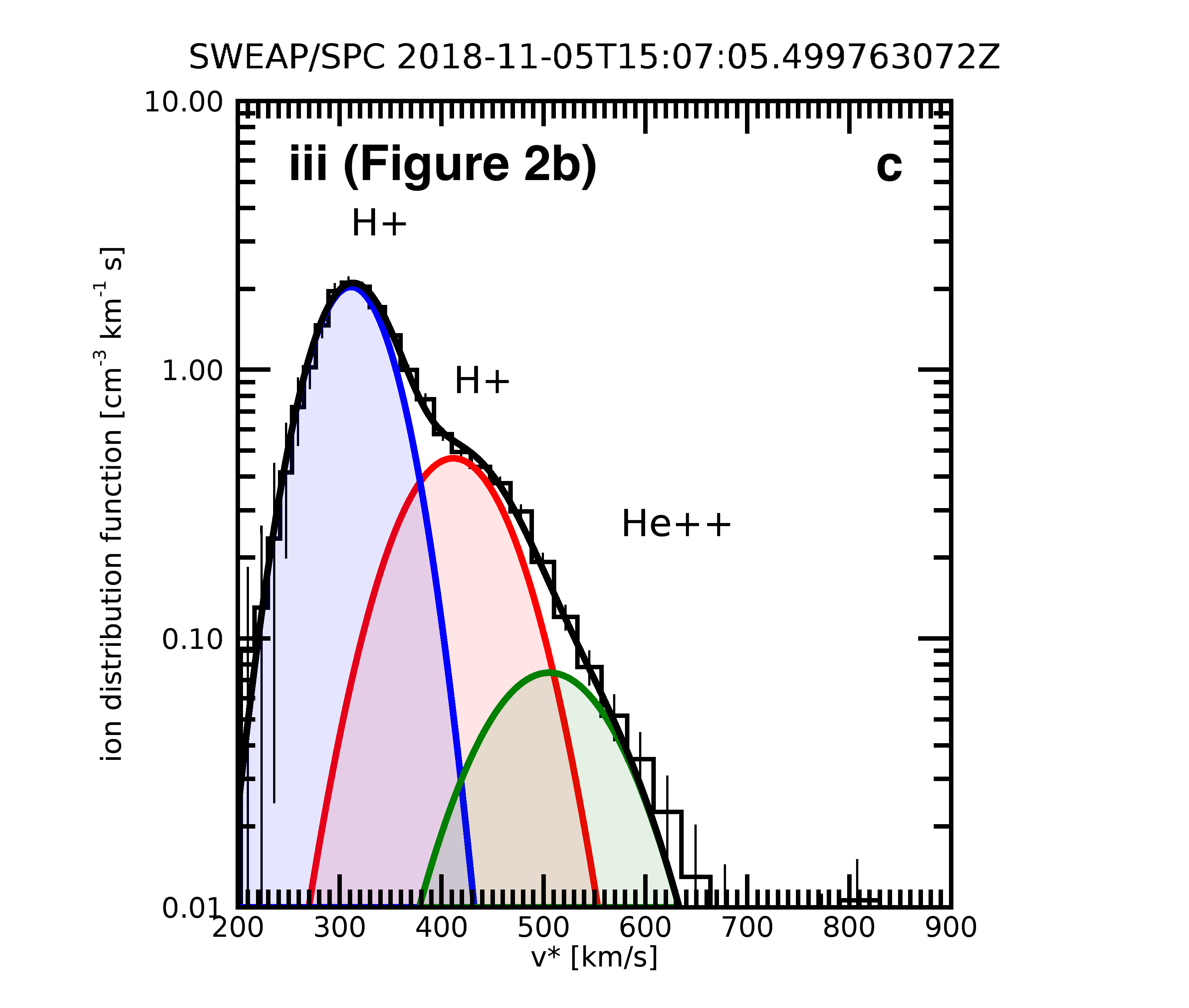}
    \end{minipage}
      \begin{minipage}{\columnwidth}
        \includegraphics[width=\columnwidth]{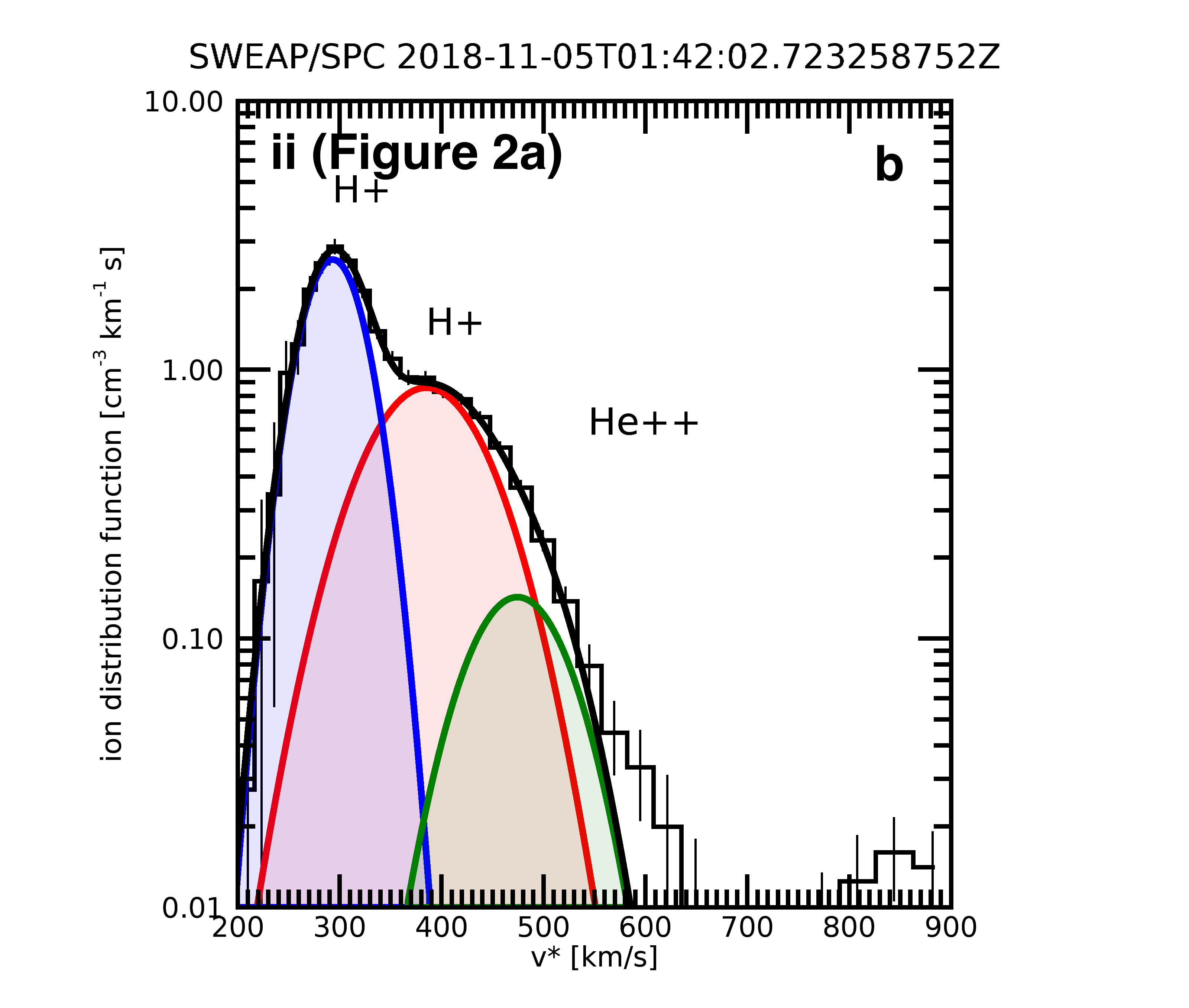}
        \includegraphics[width=\columnwidth]{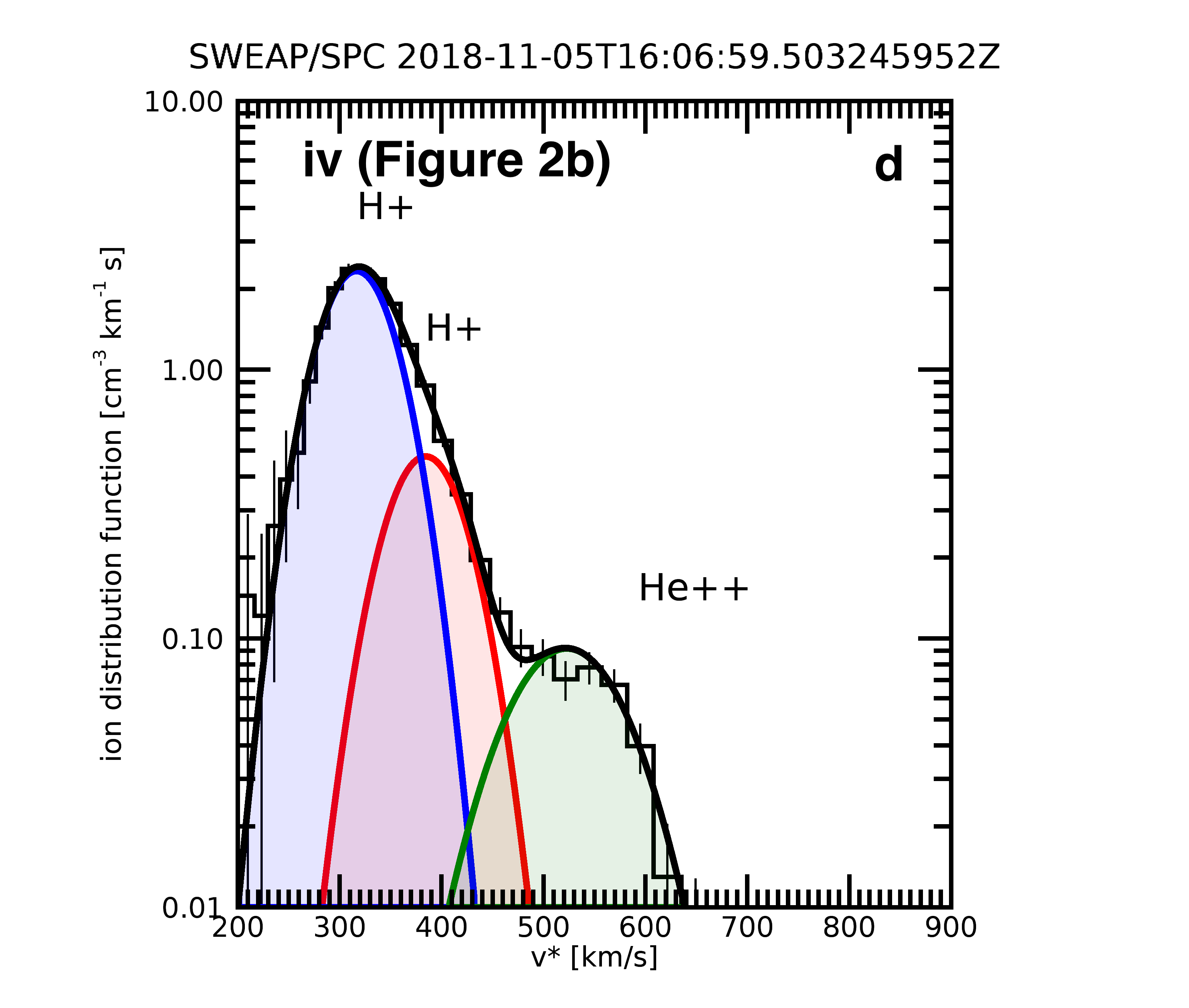}
    \end{minipage}
    \caption{(a) SPC 1-D velocity distribution function fit with two Maxwellian proton populations at time with no circular polarization shown in Figure \ref{fig:2}(a) (line i); the $\alpha$-particle population is additionally fit to a Maxwellian and is shown at the proton equivalent speed, effectively shifting the peak speed by a factor of $\sqrt{2}$. (b) SPC 1-D velocity distribution function at time with large circular polarization shown in Figure \ref{fig:2}(a) (line ii). (c) Velocity distribution function at time with no circular polarization shown in Figure \ref{fig:2}(b) (line iii). (d) Velocity distribution function at time with large circular polarization shown in Figure \ref{fig:2}(b) (line iv).}
    \label{fig:10}
\end{figure*}

\section{Acknowledgements}
The FIELDS and SWEAP experiments on the Parker Solar Probe spacecraft were designed and developed under  NASA  contract  NNN06AA01C.  The authors acknowledge the extraordinary contributions of the Parker Solar  Probe  mission  operations  and  spacecraft engineering  teams  at  the  Johns  Hopkins  University Applied Physics Laboratory. C.H.K.C. is supported by STFC Ernest Rutherford Fellowship ST/N003748/2. K.G.K. is supported by NASA ECIP Grant 80NSSC19K0912.
\appendix

\begin{figure*}
\centering
   \includegraphics{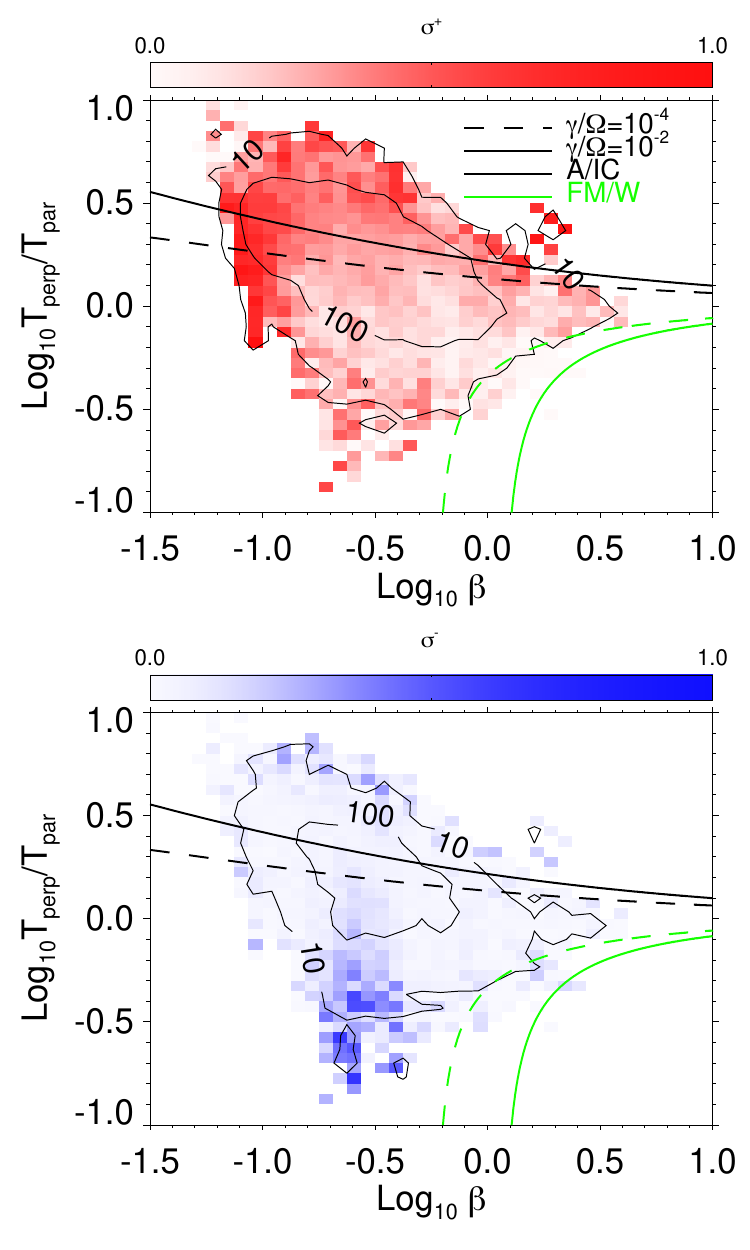}
   \caption{Distribution of encounter 1 measurements in $\beta_\parallel$-$T_\perp/T_\parallel$ plane. For each measurement the maximum values of ion and electron resonant polarization ($\sigma^\pm$) are computed. (a) Data is colored by the mean ($\sigma^+$) for each measurement showing ion resonant polarization. (b) Data is colored by the mean ($\sigma^-$) showing electron resonant polarization. Contours show 10 and 100 count levels of measured waves. Alfv\'{e}n/ion cyclotron (black) and fast-magnetosonic/whistler instability (green) thresholds are plotted at $\gamma/\Omega_p=10^{-2}$ and $\gamma/\Omega_p=10^{-4}$ levels from \cite{Verscharen2016}.}
    \label{fig:11}
\end{figure*}

\section{Reaction Wheels}\label{appendix}

\begin{figure*}
\centering
   \includegraphics{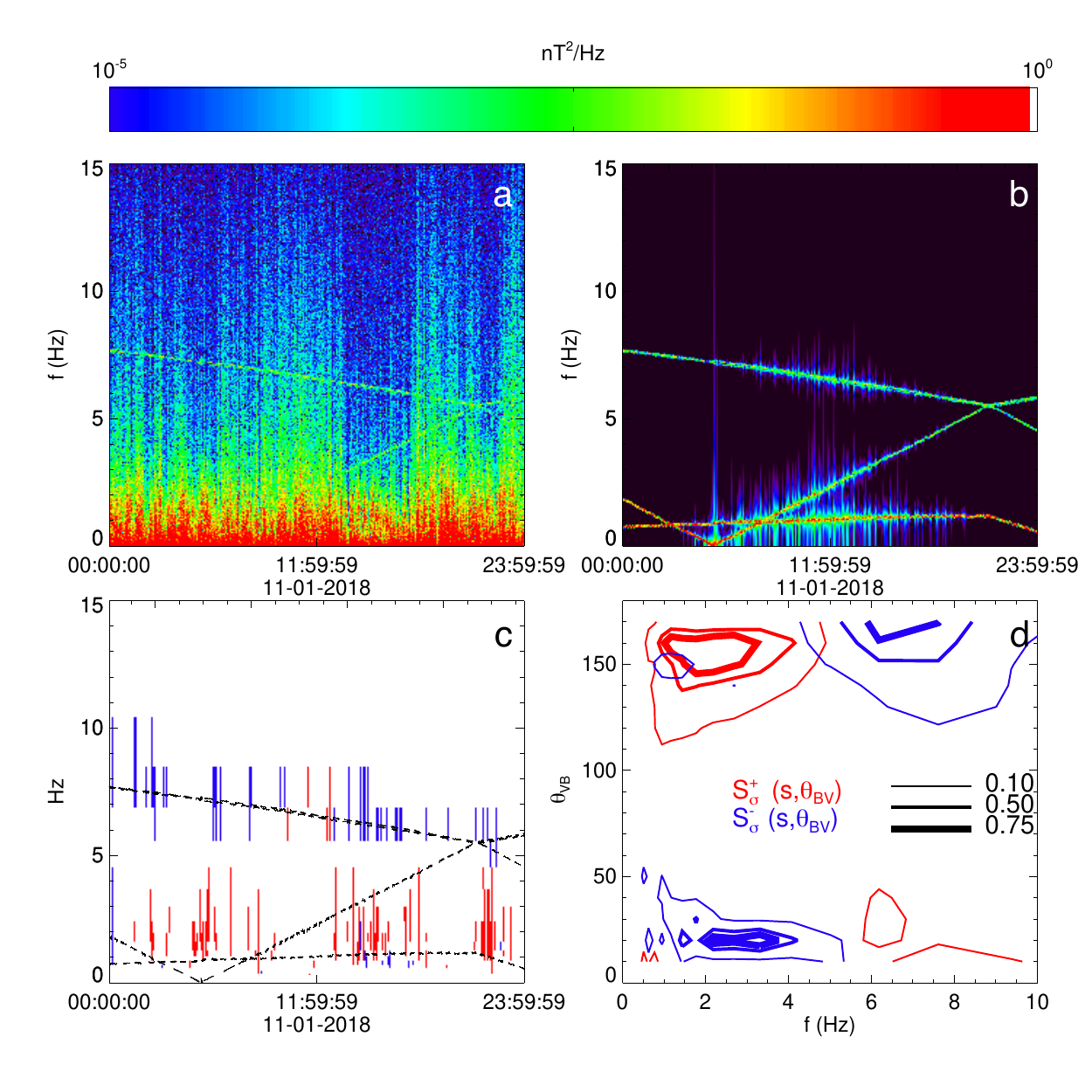}
   \caption{(a) Short time Fourier spectrogram of $x$-axis magnetic field on Nov 01, 2018. (b) Short time Fourier spectrogram of $x$-axis magnetic field on Nov 01, 2018 when retaining only frequencies contaminated by reactions wheels. (c) Intervals with negative reduced helicity corresponding to apparent electron resonant polarization (blue) and positive reduced helicity corresponding to an apparent ion resonant polarization (red). (d) Distribution of $S^\pm$ circularly polarized power in plane of $f and \thetabv$ when reaction wheel signatures are not removed; compare to Figure \ref{fig:8}.}
    \label{fig:12}
\end{figure*}

The Parker Solar Probe spacecraft is outfitted with four reaction wheels which, while necessary for maintaining stable on-orbit pointing, contribute a considerable amount of large amplitude coherent noise to the FIELDS observations. The wheels rotate at frequencies ranging from less than 1 Hz to several tens of Hz. Rotation is often coupled with multiple wheels rotating at (or very near) the same frequency and drifting together over time, though each wheel is in principle capable of rotating at a unique frequency. Each wheels generates a magnetic signature at its rotation frequency, which is easily observable in both survey and burst mode data from the FIELDS magnetometers. Though the wheels are confined to a range of rotation frequencies, harmonics and beating between wheels is observable as narrow-band noise at much higher frequencies. Additionally, autonomous spacecraft thruster firings are used to ensure that the momentum of the spacecraft stays within orbital requirements, allowing for the reaction wheels to rapidly change rotation rates without endangering spacecraft pointing. During the first encounter a single autonomous firing occurred on Nov 06, 2018 around 08:26.

The large amplitude coherent signals generated by the spacecraft reaction wheels, e.g. visible in the bottom panels of Figure \ref{fig:2},  may contaminate measurements sensitive to polarization of the environmental signal. Figure \ref{fig:12}(a) shows a spectrogram from Nov 01, 2018 computed using a short time Fourier transform; narrow-band features are observed in the spectrogram corresponding to the wheel rotation frequencies. Pre-processing and de-noising of the FIELDS data may be required in studies which are sensitive to reaction wheel signatures. While this appendix demonstrates one technique to address contamination from reaction wheels, in principle, tailored methods should be used in a case by case basis in order to minimize effects from artifacts resulting from data processing.

Reaction wheel rotation rates inherently drift in order to ensure stable pointing of the spacecraft. Typical drift rates are on order of Hz/day. However, over sufficiently short time intervals, the drift of the wheel frequencies is negligible such that electromagnetic contamination is confined to a finite, narrow-band, range of frequencies. In order to remove the reaction wheel signals, a narrow-band notch filter is used to attenuate power at each of the wheel rotation frequencies $f_{wj}$ in the Fourier domain. For an interval $N\Delta t$ the frequency resolution of the Fourier transform is $\Delta f=f_s/N$, where $f_s=1/\Delta t$. The drifting wheel frequencies are confined to a narrow range of spectral bin by choosing a sample length $N$ which confines the drifting wheel power to a single bin $\Delta f$. The rate of change of the wheel frequencies, $df_{wj}/dt$, is measured from spacecraft house keeping data. The sample size $N$ is chosen such that the maximum rate of change of the wheel frequency over $N$ corresponds roughly to the frequency resolution $\Delta f$.

$$\frac{df_w}{dt}\bigg\rvert_{max}N\Delta t=\frac{f_s}{N}$$
$$N=\sqrt{\frac{f_s}{(df_w/dt)\Delta t}}$$

For each $N$ samples, the magnetic field is Fourier transformed; coefficients corresponding to the contaminated frequencies ($f_{wj}$) are attenuated by -80 dB and the inverse Fourier transform is taken. In practice, for each day of down-sampled data has a total number of samples of $N_T=3164060$ samples, the factor of $N_T$ closest to $N$ is chosen as the sample length. For Nov 01, 2018 $N=1124$ samples (approximately $\sim 30$ s) A processed time series is then given as the reconstructed set of $M=N_T/N$ intervals.  

Figure \ref{fig:12}(b) shows the spectrogram of Nov 01,2018 when only frequencies associated with $f_{wj}$ are kept. Figure \ref{fig:12}(c) shows the indices of $|\sigma|=|<S_3/S_0>| >0.7$ computed from the wavelet transform in Section \ref{sec:3}; the reaction wheel frequencies are shown as dashed lines. Positive $\sigma$ shown in red and negative $\sigma$ shown in blue. A signature of circular polarization follows one of the drifting reaction wheels over the day between 6-8 Hz. The spectral power in these frequencies is commonly dominated by the narrow-band spectral line, suggesting that observation of polarization in these frequencies is are likely due to contamination by the reaction wheel. Around mid-day the polarization flips, however this is likely due to the definition of polarization with respect to the local mean field; this is consistent with the inversion of handedness at frequencies greater than 5 Hz is seen in \ref{fig:12}(c). 

The reaction wheel signal contributes polarized power which can be misidentified in statistical surveys of coherent power. For example, when computing the fractional power in frequency with $\thetabv$, a strong signature of negative $\sigma$ is observed at frequencies greater than 5 Hz similar to the coherent wave events. However, after processing out frequencies with reaction wheels, this feature is no longer present (e.g the Nov 01, 2018 panel shown in Figure \ref{fig:8}). 

The primary goal of this processing method is to remove narrow-band power with strong polarization signatures which contaminate the wavelet transform of the magnetic field observations. The large wavelet bandwidth (in comparison to the narrow-band wheel signature) acts to average the spectral components near the central frequency of each wavelet scale. When large amplitude coherent power from a reaction wheel is present within a wavelet's bandwidth, the wavelet response to the reaction wheel signature may dominate, resulting in a contaminated measurement. By attenuating the reaction wheel signal, the wavelet response captures the average power and phase of the non-contaminated frequencies near its central frequency.

The authors considered restricting the study to only wave events which were localized from reaction wheel signatures. In essence, an alternative approach to controlling for reaction wheels effects is to avoid any data processing, and to simply remove coherent features occurring near contaminated frequencies from the ensemble of events. However, many coherent wave events with frequencies near contaminated frequencies have enough power that the electromagnetic signature from the reaction wheel negligible--i.e. the wave rms amplitudes are commonly of order nT versus the reaction wheel amplitudes at a fraction of a nT. It is thus preferable to remove the narrow-band power at the contaminated frequencies and let the wavelet respond to the remaining power in nearby frequencies rather than simply removing the all events which have the possibility of contamination from the catalog of observed events.

The discrete joining individually processed intervals inevitably introduces artifacts in the time series: specifically, small discontinuities will exist at the edge of each processed interval. These discontinuities predominately affect the high end of the signal bandwidth near the Nyquist ($\sim18$ Hz). The fraction of power removed at each time is a small fraction of the total observed power such that discontinuities do not drastically change the structure of the magnetic field measurements. However, some measures of the magnetic field, e.g. increment based analyses commonly used to study solar wind turbulence, will certainly be sensitive to these discontinuities and alternative methods for compensating for reaction wheels will be required.

\end{document}